# Machine learning enhanced atom probe tomography analysis: a snapshot review

Yue Li [1*], Ye Wei [2*], Alaukik Saxena [1], Markus Kühbach [3], Christoph Freysoldt [1], Baptiste Gault [1,4*]

[1] Max Planck Institute for Sustainable Materials, Max-Planck-Straße 1, Düsseldorf, 40237, Germany

[2] Ecole Polytechnique Fédérale de Lausanne, School of engineering, Rte Cantonale, 1015 Lausanne, Switzerland

[3] Physics Department and Center for the Science of Materials Berlin, Humboldt-Universität zu Berlin, Zum Großen Windkanal 2, Berlin, 12489, Germany

[4] Department of Materials, Imperial College, South Kensington, London, SW7 2AZ, UK

*Corresponding authors, yue.li@mpie.de (Y. L.); ye.wei@epfl.ch (Y. W.)

b.gault@mpie.de (B. G.)

**Abstract**

Atom probe tomography (APT) is a burgeoning characterization technique that provides compositional mapping of materials in three-dimensions at near-atomic scale. Since its significant expansion in the past 30 years, we estimate that one million APT datasets have been collected, each containing millions to billions of individual ions. Their analysis and the extraction of microstructural information has largely relied upon individual users whose varied level of expertise causes clear and documented bias. Current practices hinder efficient data processing, and make challenging standardization and the deployment of data analysis workflows that would be compliant with FAIR data principles. Over the past decade, building upon the long-standing expertise of the APT community in the development of advanced data processing or "data mining" techniques, there has been a surge of novel machine learning (ML) approaches aiming for user-independence, and that are efficient, reproducible, and robust from a statistics perspective. Here, we provide a snapshot review of this rapidly evolving field. We begin with a brief introduction to APT and the nature of the APT data. This is followed by an overview of relevant ML algorithms and a comprehensive review of their applications to APT. We also discuss how ML can enable discoveries beyond human capability, offering new insights into the mechanisms within materials. Finally, we provide guidance for future directions in this domain.

# Contents

**Nomenclature and Acronyms**

| | |
|---|---|
| APT | Atom probe tomography |
| 2/3D | Two/three-dimensional |
| SDM | Spatial distribution map |
| ML | Machine learning |
| KNN | *k*-nearest neighbor |
| CSRO | Chemical short-range order |
| CMRO | Chemical medium-range order |
| CNN | Convolutional neural network |
| DT | Decision tree |
| ANN | Artificial neural network |
| DNN | Deep neural network |
| ToF-MS | Time-of-flight mass spectrometry |
| MALDI | Matrix-assisted laser desorption/ionization |
| SIMS | Secondary ion mass spectrometry |
| FEI | Field evaporation image |
| DBSCAN | Density-based spatial clustering of applications with noise |
| OPTICS | Ordering points to identify the clustering structures |
| GB | Grain boundary |
| DCOM | Distance-to-center-of-mass |
| FAIR | Findable, Accessible, Interoperable, and Reusable |
| ELN | Electronic laboratory notebook |
| LIMS | Laboratory information management system |
| RDM | Research data management |

1. Introduction

1.1. Fundamental of atom probe tomography

Atom probe tomography (APT) has emerged as a crucial characterization technique for unraveling structure-property relationships in materials with complex (micro)structures and compositions [1-3]. Compared to other microscopy and microanalysis techniques, APT occupies a unique space by combining high analytical sensitivity (10s–100s parts-per-million) with sub–nanometer spatial resolution, and APT is inherently three-dimensional (3D), which sets it apart from other techniques [3, 4]. The resolution can reach down to below 1 nm in the lateral direction and below 0.1 nm in the depth direction, depending on the considered material, the specific set of atomic planes analyzed, and the experimental conditions [5, 6]. The higher throughput and broader applicability offered by modern APT instrument designs have catalyzed an international expansion of APT since the 1990s. To date, there are over 120 APT setups in operation, that have collectively captured an estimated of at least one million APT datasets [7]. APT has now found application in engineering, geological, semiconducting, and even biological materials [2, 3, 8].

The operational principle of APT is illustrated in Fig. 1a. APT leverages the effect of an intense electric field, typically in the range of $10^{10}$ V·m$^{-1}$, generated at the tip of a specimen shaped as a sharp needle, with a radius of less than 100 nm. This electric field induces the successive desorption of each surface atom, transforming each into an ion, a process known as field evaporation [9]. Subsequently, the ion is accelerated away from the needle-shaped specimen, following trajectories set by the highly divergent distribution of the field associated with the high curvature of the specimen. These ions are then captured by a position-sensitive, time-resolved particle detector located 10–50 cm away. Precise time control over the field evaporation process is achieved through the application of short high-voltage or laser pulses [10], and this enables the determination of the elemental nature of the detected ion by time-of-flight mass spectrometry.

Fig. 1b–d outline how the ion impact position on the particle detector can be used to recalculate its original position at the surface of the specimen. Initially, in two dimensions (2D), this is accomplished using a reverse-projection model, commonly based on a pseudo-stereographic projection [11] or an azimuthal equidistant projection [12]. Through sequential processing of each detected ion, the third dimension is incrementally determined. Ultimately, APT yields a point

cloud where the distribution of each species within the material is mapped in 3D. One APT dataset typically comprises volumes of around 100×100×500 nm$^3$. For more details on specimen preparation, atom probe design, operation, and data reconstruction, we refer the interested reader to previous review articles and textbooks [2, 3, 13-16].

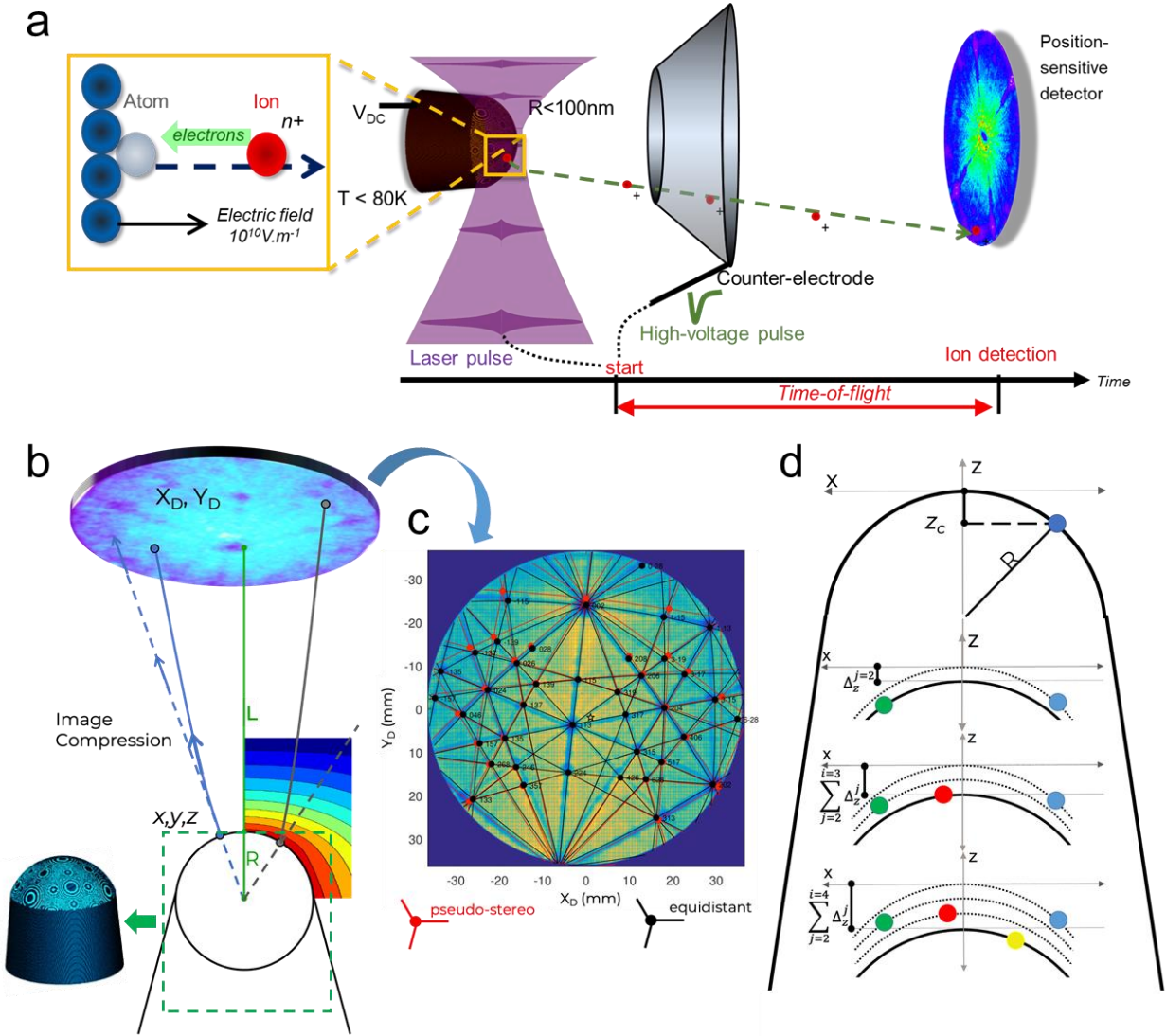

*Fig. 1 Operational principle of atom probe tomography and typical data reconstruction procedure [3, 17]. (a) Schematic of an atom probe. (b) Schematic of the ion projection from the specimen to the detector in a straight flight-path atom probe instrument. The color map illustrates a typical iso-potential estimated from 2D finite element methods calculations. (c) Comparison between projection models on an experimental pure-aluminum dataset. (d) Schematic of the procedure to reconstruct real-space depth, i.e., the z-coordinate, from top to bottom: each detected*

*ion is assumed to be projected from a hemispherical surface, and for each subsequently detected ion, an additional increment is added to the z-coordinate calculation to account for all preceding ions. Part (a) reprinted with permission from Ref. [17], CC BY 4.0. Part (b-d) reprinted with permission from Ref. [3].*

**1.2. Data format, conventional analysis workflow, and analysis challenges**

In contemporary APT instruments, a single experimental dataset can capture up to more than a billion ions. Fig. 2 illustrates the conventional data processing workflow from the raw data to the chemically resolved point cloud reconstruction [2, 18]. For each ion the raw dataset records a range of information including the X and Y hit positions on the detector ($X_{det}$ and $Y_{det}$), time-of-flight (TOF) as well as the applied standing and pulsed voltages or the laser pulse energy. From this information, the mass-to-charge-state ratio is calculated, and the X-Y-Z Cartesian coordinates of each atom can be reconstructed. As ions accumulate, peaks emerge in the mass-to-charge-state ratio histogram, forming the mass spectrum. These peak patterns correspond to the isotopes of the elements present in the material, with the peak amplitude reflecting the abundance of the specific isotope or element [19]. Due to stochastic and systematic variations in the measured time of flight, the mass-to-charge signals broaden with respect to their true values, which is affected by the instrument design, the sample's evaporation behavior, and experimental conditions. During data processing and reconstruction, APT users therefore "range" the peaks in the mass spectrum, a process where the lower and higher bounds of each mass peak is manually defined and associated with an elemental identity. This step is essential for accurately assigning elemental identities to individual data points within the reconstructed data. However, the manual nature of this process introduces variability, as the identification and ranging of peaks depend on user discretion, raising concerns about reproducibility and quantification. For instance, manual analysis of the same mass spectrum have resulted in inconsistent compositional outcomes across different users and laboratories [20, 21].

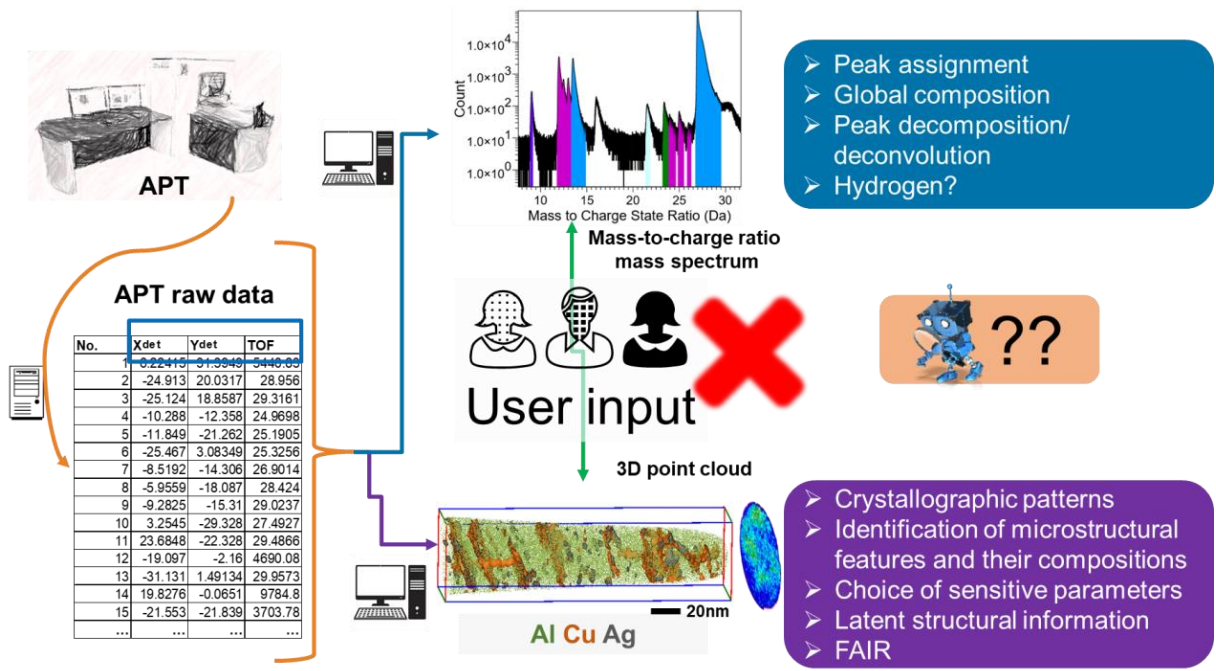

*Fig. 2 Typical APT data format and analysis workflow. From processing APT raw data to forming and analyzing the mass spectrum, and subsequently reconstructing the 3D point cloud, each step typically requires user input. This dependency underscores the potential for machine learning to enable intelligent, automated and standardized data processing workflows.*

In the analysis of crystalline materials using APT, the structure and orientation of the sample result in faceting of the specimen that introduce subtle irregularities in the electric field distribution. These result in trajectory aberrations and in the formation of patterns made of so-called poles and zone lines on the detector hit map, as in Fig. 1c [3, 22]. This pattern is characteristic of the crystalline structure and orientation and proves that partial crystallographic information is available within the APT data. Following reconstruction of the point cloud, interplanar spacing, angles between sets of planes or directions, and relative orientations between these planes with respect to the tomographic reconstruction can be extracted (Supplementary Fig. 1). Data processing techniques have been developed to quantify the periodicity corresponding to the atomic planes, typically including Fourier transform [23], 3D Hough transform [24, 25], and methods based on radial distribution functions [26] including spatial distribution maps (SDMs) [27, 28]. This partial structural information is critical for calibrating data reconstruction and characterizing orientation relationships between microstructural features [29, 30]. Until now, crystallographic information, including the identification of the sets of atomic planes, their orientations, and 3D orientation mapping, is typically extracted manually, relying on user interpretation. This manual approach is error-prone and underutilizes much of the inherent crystallographic information, highlighting the need for automated methods to extract and analyze this data.

Once the specimen's tomographic reconstruction is generated, analytical tools can be deployed to analyze the spatial distribution of chemical species. As exemplified in Supplementary Fig. 2, the iso-surface analysis, based on composition or point density, highlights interfaces, such as precipitate-matrix and grain boundaries [31-34]. The *k*-nearest neighbor (KNN) analysis reveals elemental segregation tendencies through a leftward/rightward shift in the peak in the KNN distance distribution [35]. KNN distance can then be used to segment populations of atomic clusters out of the data for instance [36, 37]. Radial distribution functions also capture solute interactions and allow for quantification of precipitate populations for instance [38, 39]. Composition profiles and proximity histograms quantify segregation as a function of the distance either to iso-surfaces or along a specific region-of-interest [40, 41]. These analyses quantify the local composition that can be correlated with microstructural information.

Determining critically-sensitive parameters, such as the concentration threshold for iso-surface-based methods or the maximum separation distance in cluster-finding algorithms, remains challenging [42-44]. This can introduce inconsistencies across users and compromise the robustness of feature identification. Moreover, many local and subtle features in APT datasets often go unnoticed due to the anisotropic spatial resolutions and limited detection efficiency of the technique [6, 45, 46]. For example, latent structural information with optimal depth resolution near crystallographic poles has yet to be fully exploited for analyzing local chemical ordering [5, 6]. The human-centered analytical procedures often introduce significant issues and biases, which hinder efficient data processing and limit the ability to make the vast amounts of APT datasets findable, accessible, interoperable, and reusable (FAIR) [20, 47, 48].

As one of the fastest-growing scientific fields, machine learning (ML) holds immense promise for processing and analyzing APT data. Over the past decade, advanced ML approaches have been increasingly applied within the APT community to tackle challenges such as automated mass spectrometry analysis, intelligent extraction of crystallographic information, and efficient identification of intricate microstructural features, all in line with the FAIR principles. These innovations have led to the development of novel methods aimed at achieving user independence, efficiency, reproducibility, and robustness. ML empowers users to analyze APT data effectively, regardless of their expertise, while mitigating human bias and uncovering hidden patterns within complex datasets. This capability not only standardizes data interpretation but also reveals new insights into the physical mechanisms underlying material behavior. Despite these advancements, a comprehensive summary of ML-enhanced APT techniques is still lacking, and there is a pressing need to outline future development directions in this emerging field. In this review, we present a rapid overview of the current state of ML in APT, including a survey of relevant ML algorithms, a detailed examination of recent applications, and a forward-looking perspective on potential advancements in ML-enhanced APT direction.

## 2. Relevant machine learning algorithms

3D data point clouds defined by their x, y, and z coordinates and a corresponding fourth dimension, namely the mass-to-charge ratio in APT, are commonly used for representing and analyzing the geometrical or spatial distribution of objects or properties. Substantial progress has been made in processing and analyzing 3D point clouds, utilizing a range of ML-based algorithms for tasks such as data classification and data segmentation [49]. In this section, we provide an introduction on the ML in general terms. ML is essentially a computer algorithm that aims to learn the mapping (the rules) that describe the relationship between a set of inputs and outputs. The model then allows to make predictions based on this mapping (and rules) learned iteratively from the set of data used to train the model. ML is widely employed in a range of computing tasks in which explicit programming is practically impossible [50, 51]. Broadly speaking, there are two type of ML techniques that can be applied to APT research: supervised learning and self-supervised learning.

### 2.1. Supervised learning

Supervised learning involves training a model using a dataset of N labeled examples $(x_1, y_1)$, ..., $(x_N, y_N)$, where each $x_i$ represents the input features of the $i$-th example, and $y_i$ is its corresponding label, either a class or a continuous value. The goal is to learn a function $f: X \rightarrow Y$, mapping inputs $X$ to outputs $Y$, from a set of possible functions $F$. The objective is to approximate $f$ so that the model can make accurate predictions on new, unseen data. To achieve this, a loss function is used to measure the difference between the model's predictions and the true labels. The loss quantifies how "far off" the model's predictions are and guides the training process. Minimizing this loss helps the model assign inputs to correct outputs with the highest possible accuracy and confidence. Let us take an example where supervised learning can be applied in APT. Each atom in the dataset needs to be assigned to a specific elemental identity before further processing. Because peaks exhibit patterns associated to the elements' isotopic abundance distribution that can be learnt, supervised learning is a natural solution to this problem. Supervised learning techniques generally fall into one of the following categories:

- Classification, where the outputs are discrete labels.
- Regression, where the outputs are continuous and real-valued.

In the classification task, the data is categorized into two or more classes. The ML model needs to learn to assign inputs to one or more of these classes. In other words, the model needs to learn a set of 'decision boundaries' that allows to discriminate the different classes while minimizing the error. Regression is similar to classification, however, the outputs are continuous rather than discrete.

A decision tree (DT) is a flowchart-like learning model that partitions the data into individual subgroups based on learned rules [52]. DT resembles human decision making, which leads to a straightforward model interpretation [53]. Branching in a tree is based on conditions (control statements or values). The data points are split into either side of the node, depending on whether it meets the specific conditions (Fig. 3a). A DT can be particularly suited for analyzing extensive datasets, due to its fast inference speed [54]. The DT-based methods have been used to analyze a wide range of time-of-flight mass spectrometry (ToF-MS) spectra (Section 3.1).

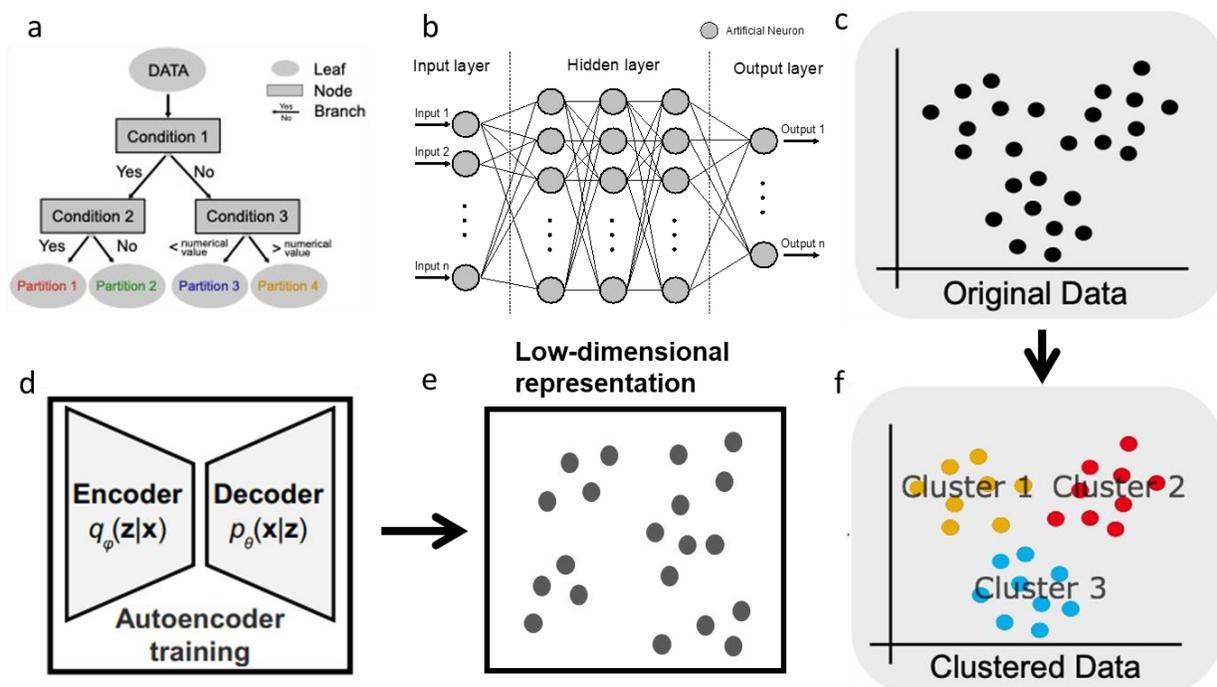

*Fig. 3. Typical machine learning algorithms used for APT learning. (a) A decision tree based on conditions learned from the data. (b) A deep neural network that consists of Input, Hidden and Output layer. (c and f) Clustering tasks (d and e) Autoencoder learns the data representation at a reduced dimension.*

An artificial neural network (ANN) is a machine learning model inspired by the structure of biological neural networks in the human brain. It consists of multiple interconnected artificial neurons, each linked by weighted connections, forming a layered network structure [55]. A deep neural network (DNN) extends the concept of an ANN by stacking multiple layers of artificial neurons, typically comprising an input layer, one or more hidden layers, and an output layer, as illustrated in Fig. 3b [56]. The ANN-based methods have been applied to identify atom probe crystallographic characteristics of aluminum (Section 3.2) and to extract local chemical ordering structures from the matrix in an array of alloys (Section 3.3).

**2.2. Self-supervised learning**

In the scenario of self-supervised learning, data without labels is given. It is the task of the model to find interesting and complex structures hidden within unlabeled data. This is particularly interesting for APT data as it contains billions of atoms without any labels [57]. Two of the most important self-supervised learning techniques are:

1. Clustering - data is divided into separate subgroups called clusters.
2. Dimensionality reduction - data is mapped into a lower-dimensional space.

In clustering, a set of data points is divided into clusters in such a way that data in the same cluster are more "similar" to each other than to those in other clusters (Fig. 3c, f). Such "data clusters" must not be confused with "solute clusters" in materials, i.e., when minority species are found in close spatial proximity rather than being randomly distributed. Similarity of data points is measured by means of a defined set of criteria e.g., the Euclidean distance between two data points. Unlike supervised learning, information about the clusters is not specified beforehand. Therefore, cluster-finding algorithms are archetypical self-supervised tasks, and analyzing a population of solute clusters is a common and important tasks in APT data processing [58].

The dimensionality of a dataset is the number of features or variables it contains, each representing a distinct attribute that describes the data points. APT data inherently contains many variables originating from different sources, such as signals from the particle detectors, time-of-flight mass spectra etc. A large numbers of variables generally leads to a sparse representation of data. Due to this sparsity, the ML models have a strong tendency towards overfitting on the APT data [59]. In other words, the model might perform exceptionally well on a specific type of materials data but fails to generalize effectively to other datasets. This is often referred to as the "curse of dimensionality". It can thus be desirable to transform the variables into a new set with lower dimensionality, while not losing much information. This transformation is often referred to as "dimensionality reduction" [60]. Dimensionality reduction is mainly used to remove correlated variables and redundant features. One of the widely used technique is the autoencoder, which consists of two main parts: an encoder, which compresses the input data into a lower-dimensional representation, and a decoder, which reconstructs the original data from this compressed representation (Fig. 3d) [61]. The goal is to train the network to learn a compressed representation of APT data, making it easier to reveal material phenomena by analyzing these low-dimensional representations (Fig. 3e).

## 2.3. 3D-point-cloud-based algorithms

There are typical 3D point cloud learning techniques that have potential for application to APT data including:

1. **Segmentation in 3D**: Dividing a point cloud into meaningful subsets, which are often used to isolate objects or features [62]. This can be used for separation of specific region of interest in the APT data in a automatic manner.

2. **Feature extraction**: Identifying and extracting significant features from the point cloud, such as edges, corners, and keypoints (e.g., Harris 3D keypoint detection [63]). This task is of particular importance, materials scientists often have akin interest in atoms or clusters with specific features, which could be attributed to some known or unknown mechanisms.

3. **Object recognition**: Detecting and classifying objects within the point cloud, leveraging machine learning algorithms to enhance accuracy and robustness [49], as the current protocol involves extensive human labelling of objects appeared in the APT data

## 3. Application of machine learning to APT

In this section, we provide a detailed review of advancements in mining APT data using ML, structured according to the typical APT data analysis workflow illustrated in Fig. 2. Accordingly, this section begins by exploring ML-enhanced techniques for mass spectrometry analysis. It then transitions to methods for intelligently extracting crystallographic information from detector hit maps, which is vital for calibrating reconstruction parameters. Following the reconstruction of the 3D point cloud, ML facilitates the efficient discovery of the rich microstructural features inherent in APT datasets, surpassing human capabilities. These features include local chemical ordering, as well as insights into grain and phase boundaries. Finally, the analysis of complex APT data must align with FAIR data principles, and we detail the latest progress in this area.

### 3.1. Mass spectrometry analysis assisted by machine learning

ToF-MS is a widely applied analytical technique in mass spectrometry that determines the mass-to-charge ratio of ions by measuring their time-of-flight through an electric field [64]. This approach allows for highly precise, quantitative analysis across a broad spectrum of atomic and molecular masses [65]. ToF-MS serves as a foundational platform in various ionization techniques, including matrix-assisted laser desorption/ionization (MALDI), secondary ion mass spectrometry (SIMS), and APT, each employing unique mechanisms to ionize samples. This versatility has expanded the applicability of ToF-MS, enabling it to support a wide array of studies—from monitoring chemical reactions and characterizing large biomolecules to quantifying dopants in semiconductor materials and assessing atomic-scale impurity distributions at grain boundaries in metallic alloys [66-68].

Pioneering studies have demonstrated the feasibility of applying statistical and ML techniques to the analysis of ToF-MS spectra. For instance, self-supervised ML methods have been used for exploratory data analysis in ToF-SIMS and ToF-MALDI, aiding in pattern recognition and data interpretation without requiring labeled datasets [69-72]. More recently, Bayesian approaches have been adopted for peak identification within APT, offering a robust probabilistic framework for spectrum analysis [73, 74]. In particular, the method developed by Mikhalychev et al. [74] facilitates simultaneous identification and deconvolution of diverse ToF-APT mass spectra, significantly enhancing analytical capabilities for complex spectral environments. Fig. 4a illustrates the Bayesian methodology for ranking candidate components within a sample. In the example shown in Fig. 4a, three components A, B, and C are considered, with the assumption that only these three ions can be present in the measured spectrum. To estimate the probabilities of the components being part of the correct sample model, all possible models comprising these ions (i.e., A, B, C, AB, AC, BC, and ABC) are constructed. The posterior probabilities, representing the probability of a certain Model after observing new Data, p(Model|Data), are used to quantify the similarity between each model and the observed spectrum. It should be noted that these posterior probabilities are assigned arbitrarily for illustrative purposes, and their sum equals 100%. The posterior probabilities of individual ions, representing the probability of a certain Ion after observing new Data, p(Ion|Data), are subsequently calculated by summing the probabilities of all models that include the respective ion. For simplicity, it is assumed that the abundance of each component remains constant and equal across all models containing that component (Fig. 4a). With reasonable prior information, this method can lead to very robust results. However, prior knowledge must often be provided by users. If an inaccurate prior is assumed, the computation can become very expensive and error prone.

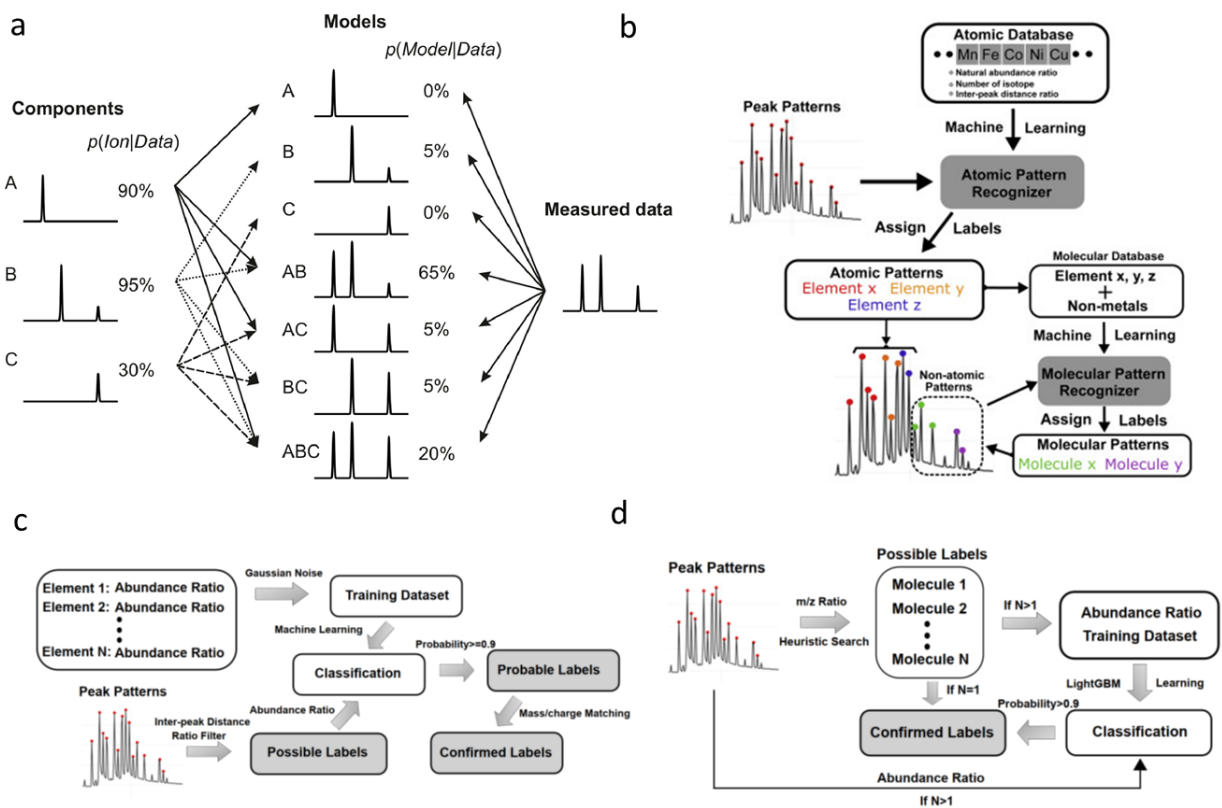

*Fig. 4. Time of flight spectrometry analysis using machine learning* [19, 74]*. (a) Bayesian approach for mass spectra peak identification. (b) Flowchart of machine learning approach for mass spectra peak identification. (c) The atomic pattern recognizer (d) on-the-fly molecular pattern recognition. Reprinted with permissions from Refs. [19, 74].*

Another approach to this problem has been attempted using decision tree-based models, called ML-ToF [19]. This approach automates the process of assigning elemental and molecular identities to peaks and series of peaks within ToF-MS spectra (Fig. 4b). Firstly, this method introduces an atomic pattern recognition approach for identifying ToF mass spectra of atomic species, as illustrated in Fig. 4c. The method relies on a database containing isotopic abundances for all elements, which is then used to train a decision tree model for classification purposes (Fig. 4c). Furthermore, the model incorporates uncertainty analysis to evaluate the extent to which peak patterns are affected by noise levels and shape features. Secondly, a molecular pattern recognition approach (Fig. 4d) is employed to address scenarios where two or more elements with different natural abundance ratios combine, resulting in a new fingerprint. This new fingerprint is distinct from the original atomic species not only in terms of atomic number but also in abundance ratios. Such combinations are frequently observed among non-metal elements (e.g., carbon, oxygen, and nitrogen) and occasionally involve metallic elements as well. This presents a significant challenge for database construction, as it is impractical to search for all possible combinations through brute force. The molecular pattern recognizer dynamically generates a training database based on the classifications made by the atomic pattern recognizer. This training database is then used to train a new classifier capable of identifying molecular patterns. With the help of this technique, the ML-ToF method has demonstrated its effectiveness in analyzing a wide range of ToF-MS spectra without requiring prior knowledge of the sample's composition, and it has been successfully applied to various material systems and analytical techniques.

## 3.2. Atom probe crystallographic analysis assisted by machine learning

Crystallographic information within atom probe datasets includes observing atomic planes, calibrating tomographic reconstruction, determining orientations and orientation relationships between microstructural features, and performing three-dimensional orientation mapping [22]. Currently, the extraction of crystallographic information is performed manually, with the identification and interpretation of crystallographic features dependent on the individual user. This process introduces a high potential for error. As a result, the full crystallographic complexity inherent in many atom probe data sets remains largely underutilized. Therefore, several methods have been proposed to automate the extraction of crystallographic information from APT data.

Fig. 5a is a schematic representation of a classic multi-metric approach used to extract latent zone line and pole information from atom probe experiments. By creating a 2D histogram of the X and Y coordinates from a sequence of detector hits on an analyzed crystalline specimen, referred to as the field evaporation image (FEI) or detector hit map, variations in point density can be observed under suitable experimental conditions. The point density serves as the primary metric for revealing zone line and pole information. The pattern arises from trajectory aberrations caused by variations in the electric field across the emitter's surface [75]. These variations result from lattice terracing and the associated local magnification effects, which are particularly pronounced at the terrace edges where electric field gradients are strongest [76-78]. Other metrics associated with each detected event, such as multiple hits, evaporation sequence, and the number of pulses between detection events (as shown in Fig. 5a), also correlate with crystallographic information. These metrics provide a powerful means to filter and further refine latent crystallographic details. These raw detector data needed to perform such analyses can be accessed through the .epos file or the .apt file, in the case of the currently available commercial instruments [78].

Another common approach for extracting crystallographic information from FEI is the use of the Hough transform to detect zone lines on the detector hit map. This approach was adapted from the so-called "orientation imaging microscopy" framework developed for electron backscattered diffraction [79, 80]. The Hough transform has also been extended in 3D for plane detection in reconstructed APT data [24, 25]. However, due to the distinct nature of the projection in APT [12], many detector hit maps do not display clear zone lines, limiting the applicability of Hough-transform-based methods for analyzing APT data (Fig. 5b demonstrates a typical example). Fig. 5c shows a machine learning approach, called ML-APX [81], which can directly process information from the positions and angular relationships between the poles within the FEI. This method employs a deep neural network to automatically compute the crystal orientation of specimens. Specifically, it starts with a noise removal on the FEI and pole detection algorithm which locates the positions of the poles. Afterwards, four poles are selected based on the minimum area and minimum angle of the quadrilateral formed between the poles. The four angles of the quadrilateral are then used as the input for the deep neural network. In this way, ML-APX can recover crystallographic information from the relative position between poles.

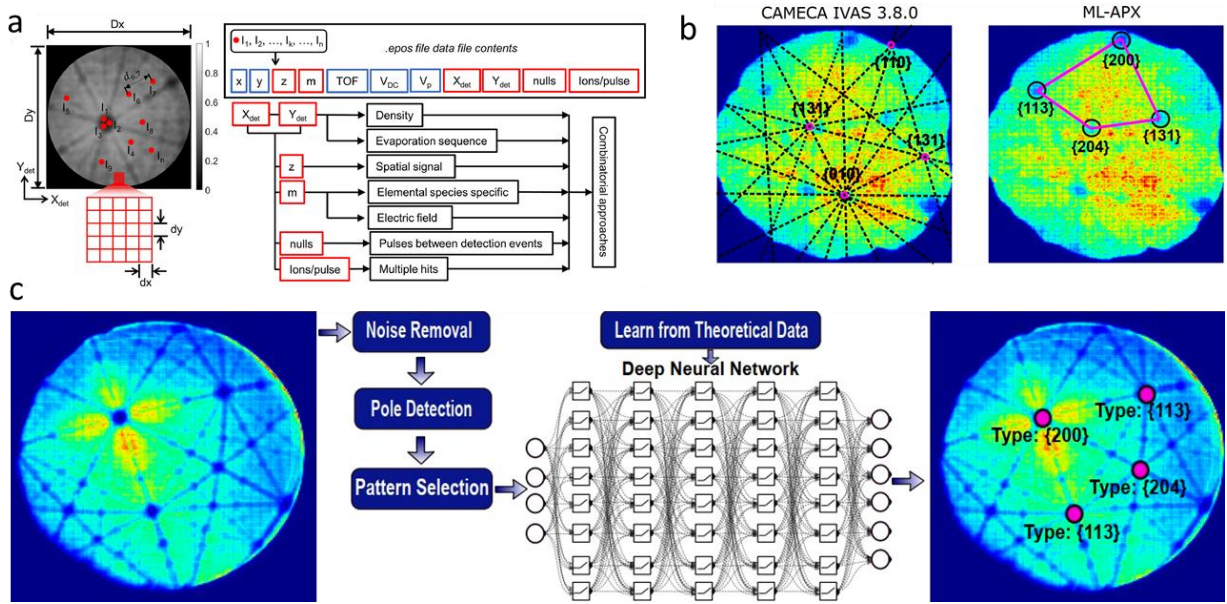

*Fig. 5. Various automatic tool for crystallographic information extraction [78, 81]. (a) A multi-metric approach has been employed to uncover latent crystallographic information within atom probe detector maps. The red dots denote individual ions impacting the detector ($I_1$, $I_2$, …, $I_i$, …, $I_n$) in the field evaporation imaging process. The underlying grey image represents a density map (or 2D histogram) of ion hits, which serves as the most common visualization for identifying poles and zone lines. The .epos file format used for these analyses contains 11 data categories for each ion hit, like reconstructed x, y, z coordinates and applied pulsed voltage $V_P$. Additionally, a data hierarchy illustrates how various data categories are employed to map specific metrics that correlate with crystallographic patterns observed across the detector. (b) The Hough-transformation based method (from CAMECA IVAS 3.8.0) v.s. ML method (ML-APX) in the presence of strong noise. While the Hough transformation assigns wrong crystallographic orientation to the poles (e.g., {131} instead of {204}) due blurred line patterns, ML-APX correctly identifies the orientation of four detected poles. (c) The ML-APX pipeline, which consists of noise removal, pole detection, pattern selection and recognition by a deep neural network that learns the crystallographic information from simulated data. Reprinted with permissions from Refs. [78, 81].*

## 3.3. Chemical ordering analysis enhanced by machine learning

Chemical ordering refers to the preferential occupation of specific sites within crystalline materials by particular elements [82]. It can be classified into long-range order (e.g., ordered intermetallic precipitates) and local chemical order/solute-atom clustering based on the range and degree of elemental rearrangement [5, 83, 84]. The occurrence of chemical ordering within a disordered crystalline matrix has been reported to significantly influence the mechanical and functional properties of materials [85-87].

### 3.3.1. Conventional algorithms

Many traditional statistical algorithms have significantly contributed to analyzing the distribution of chemical species in APT datasets. For example, iso-surface methods [31-34] were widely used to identify precipitates by highlighting regions with compositions markedly different from the surrounding matrix. Proximity histograms [41] further enable the calculation of compositional gradients on either side of these interfaces, with the caveat that the precision will be dependent on the voxelization itself [88]. To analyze smaller-scale domains, cluster-finding algorithms have been extensively used to detect nanoscale solute clusters. Techniques, such as statistical KNN analysis [35], related maximum separation [89], and core-linkage algorithm [90], rely on criteria like solute-solute distance or local composition. Other clustering algorithms have been used to segment out microstructural features based on the differences in atomic densities in APT data, for example, density-based spatial clustering of applications with noise (DBSCAN) [90], hierarchical DBSCAN [91], and ordering points to identify the clustering structures (OPTICS) [92]. Alternative approaches based on radial distribution functions [26, 38], have been proposed, to investigate ordering reactions [38] and to characterize populations of precipitates by using the mathematical framework initially developed for processing small-angle scattering data [39].

Applying clustering algorithms directly to the atomic positions in APT data can be computationally expensive, and each algorithm involves various hyperparameters that are seldom examined in a systematic manner. Currently, the most widely used approach for cluster detection is the maximum separation algorithm, a heuristic method heavily dependent on user-defined parameters [89, 90]. The Gaussian mixture model [93], however, offers a higher level of accuracy by probabilistically assigning atoms to clusters. This approach effectively captures scientifically meaningful uncertainty, especially for atoms near precipitate/matrix interfaces, enabling a more refined interpretation of clustering behavior.

While these methods have proven valuable, they also have limitations. One notable issue is the lack of consistency in defining parameters to differentiate clusters from the matrix [42]. For instance, the determination of the maximum distance $d_{max}$ in the maximum separation algorithm can be significantly affected by local density variations resulting from trajectory aberrations, potentially resulting in inaccurate results. Moreover, using APT data to identify fine domains with diameters below approximately 2 nm (such as chemical short-range order (CSRO) with a diameter of about 1 nm, depending on the material) remains challenging due to anisotropic spatial resolutions and limited detection efficiency [6, 45, 82, 94].

### 3.3.2. Deep learning enhanced analysis of chemical short-range ordering

SRO is commonly characterized using the Warren–Cowley parameter and its derivatives [46, 95, 96]. These SRO parameters [95, 97] are specifically designed to reveal the $1^{st}$ NN and $2^{nd}$ NN chemistry for analyzing CSRO within APT datasets. However, caution is necessary when employing this approach for CSRO analysis due to the inherent anisotropic spatial resolutions and imperfect detection efficiency of APT. To address these limitations, He et al. [96] recently introduced a data-driven approach to measure SRO using APT in a CoCrNi alloy. This method compensates for APT's constraints by leveraging threshold values of SRO to delineate regions where sufficient atomistic neighborhood information is retained versus regions where it is not. By estimating the extent of spatial resolution errors and detection inefficiencies, researchers can apply correction factors to approximate the true SRO, enhancing the reliability of their analyses.

Li et al. [5, 82] chose a different approach, whereby a series of investigations employing 1D/2D convolutional neural network (CNN) enabled tomographic imaging of medium/short-range orders in various engineering materials by APT. Their proposed strategy, referred to as ML-APT and illustrated in Fig. 6, leverages the highest quality, near-atomic depth resolution of APT along with its high elemental analytical capability based on the SDMs along the depth direction. The core workflow involves several steps. First, a simulated library of SDMs is created (Fig. 6a) and used to train CNNs to develop a disordered/ordered structure recognition model (Fig. 6b). Then, experimental SDMs are calculated from APT datasets of alloys where CSRO is known or suspected to significantly influence properties. Finally, experimental SDMs are analyzed using the recognition model to uncover both chemical and crystallographic information related to CSRO.

This method was applied to address the long-standing K-state behavior in body-centered cubic Fe-Al and Fe-Ga alloys [82, 98]. Following validation against artificial data for ground truth, ML-APT was applied to reveal non-statistical B2-CSRO (FeAl) instead of the generally-expected $D0_3$-CSRO ($Fe_3Al$) in Fe-Al alloys and modified $D0_3$-CSRO ($Fe_3Ga$, or $L6_0$ structure) in Fe-Ga alloys (Fig. 6c). Quantitative information was also extracted (Fig. 6d). However, ML-APT necessitates prior knowledge of possible CSRO configurations, such as $D0_3$ and B2-like structures in Fe-Al.

For medium/high-entropy alloys where potential CSRO configurations are unknown in advance, Li et al. [99] further proposed an extended ML-APT strategy to quantify the CSRO domains in 3D APT data from CoCrNi medium-entropy alloys without prior CSRO configuration knowledge. By obtaining all information from pairs of the same species along different crystallographic poles, the $L1_2/DO_{22}$- and $L1_1$-CSRO configurations were ultimately deduced, along with their elemental site occupancy and size/morphology distributions (Supplementary Fig. 3). With the assistance of ML-APT, the size limitation of recognized CSRO domains in APT data has been pushed to below 1 nm, depending on the alloy system and the quality of experimental data.

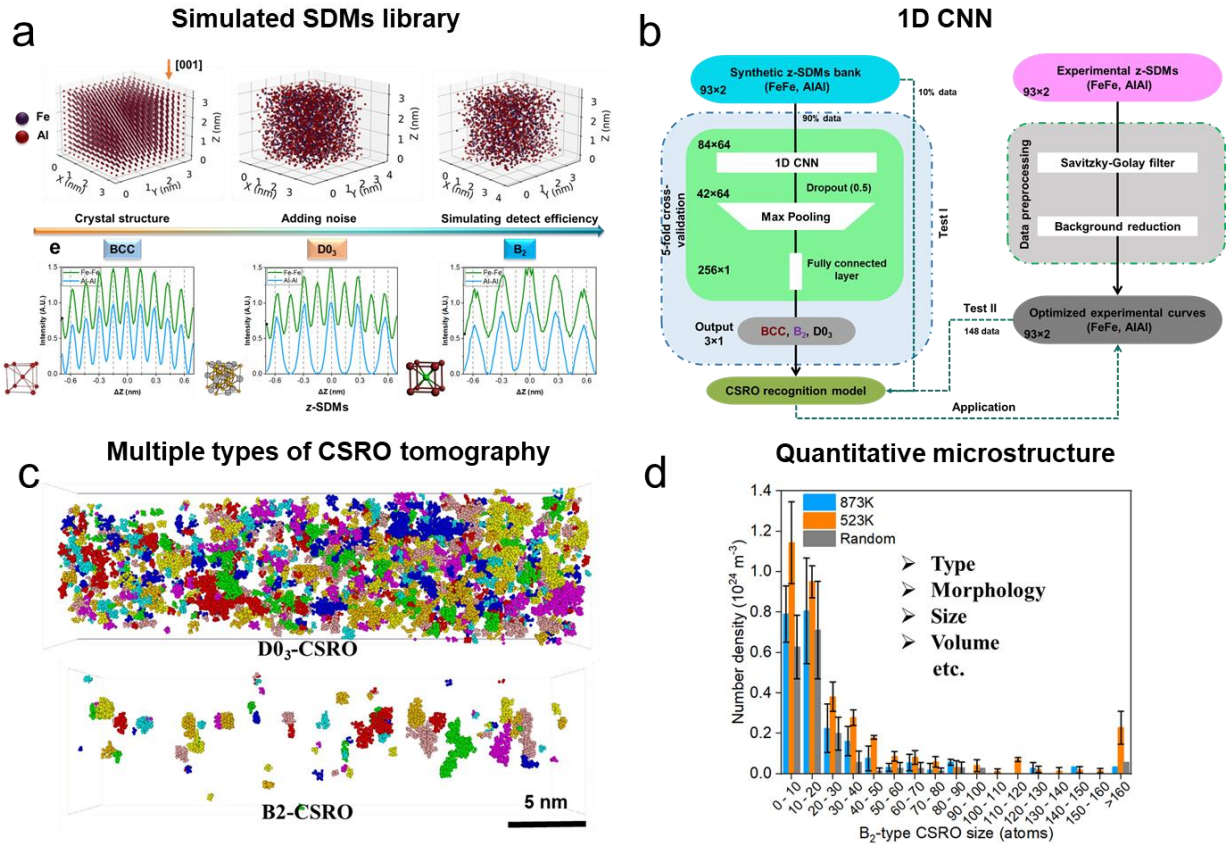

*Fig. 6 Machine-learning-enhanced APT strategy to find CSRO in Fe-Al/Ga alloys [82]. (a) Procedure for constructing multiple crystal structure classes and generating corresponding synthetic z-SDMs. (b) Proposed 1D CNN framework to find CSRO within APT data. (c) Recognized two types of 3D CSROs. (d) Quantitative CSRO information, such as type, morpholodgy, and size distributions. Reprinted with permission from Ref. [82], CC BY 4.0.*

### 3.3.3. 3D deep learning enhanced analysis of chemical ordering

Although ML-APT demonstrated remarkable classification accuracy in resolving CSRO, the conversion of the 3D periodic arrangement of atoms into 2D/1D SDMs could result in information loss and decrease tolerance to offsets in atomic positions caused by APT resolution limits and missing atoms from the detection efficiency. To address this, a machine learning-based approach based on 3D CNNs was proposed, directly operating on the original periodic arrangements of atoms [100]. This approach was applied to unveil the 3D distribution of $L1_2$-type $Al_3(LiMg)$ precipitates with an average radius of approximately 2.5 nm in an FCC Al-Li-Mg system, consistent with segmentation based on typical iso-surface methods [5, 32, 100]. However, 1D/2D/3D CNN-based methods necessitate voxelization, integrating signals from a specific volume of atoms (1–2 nm), limiting the size range of recognized domains using CNN and not reaching down to the single-atom scale.

To address this limitation, Yu et al. [101] further proposed AtomNet, a 3D point-cloud-based neural network capable of handling information at the single-atom level without voxelization, thereby revealing various nanoscale microstructures. AtomNet considers both compositional and structural information, enabling the recognition of different microstructures at the single-atom level ranging from $Al_3(LiMg)$ δ′ precipitates in an Al-Li-Mg alloy to chemical medium-range orders (CMROs) in Au-Cu, and even 2D stacking faults in Co-based superalloy. AtomNet surpasses the above-mentioned iso-surface and 2D CNN-based methods [5, 32] by its ability to detect nano-domains independently of crystallographic poles and reveal small CMROs without evident elemental segregation. Moreover, AtomNet can display unseen structures not present in the training data, such as stacking faults in a Co-based superalloy because of the local difference in the local atomic environment. However, while AtomNet significantly improves CMRO detection, its performance in identifying smaller CSROs is less satisfactory due to interference from neighboring atoms at lower resolutions. This limitation highlights the need for more advanced 3D models capable of intelligently extracting high-quality, near-atomic depth resolution signals to analyze CSRO with expert-level precision.

### 3.4. Detection and segmentation of microstructural features enhanced by machine learning

Beyond its ability to provide detailed chemical analysis at the near-atomic scale, the value of APT lies in revealing also larger patterns in the chemical composition in 3D, patterns that link to crystallographic defects, microstructural features including different phases for instance, on length scales that span 1–100 nm. In combination, these features directly or indirectly influence the structural or functional properties of the material under investigation. These patterns can appear in a variety of shapes and appearances: precipitates in spherical, ellipsoidal, plate-like, or other geometric forms, linear or planar segregation at or near grain boundaries, stacking faults, and dislocations, as well as interfaces between grains of coexisting bulk phases, each with a distinct chemical fingerprint. ML-based techniques can aid in identifying such features directly, or segmenting the APT point cloud into coherent domains according to contrast in composition. This first step must be followed by algorithms (ML or others) to find suitable shape descriptors to capture the essential geometric character of each of these features/domains, and ideally also help preparing subsequent geometry-adapted further analysis (such as properly defined regions of interest across interfaces for proximity histograms). While much of this can be done by humans and conventional tools (notably iso-surfaces), ML offers the chance to replace human bias in choosing parameters by reproducible schemes, and thus also to scale up analysis across multiple data sets in a standardized way.

### 3.4.1. Automated identification and quantification of grain boundaries

APT can quantify chemical characteristics within grain boundaries (GBs) at a near-atomic scale. The basic idea used to capture GBs and interphases automatically is to find flat regions with enhanced solute concentration compared to the surrounding regions. Building their algorithm on this idea and using computational geometry techniques, including Voronoi tessellation and distance-to-center-of-mass (DCOM) calculations, Felfer et al. [102] could extract GBs and interphases and compositionally analyze them. DCOM detects microstructural features by analyzing solute density gradients, while Voronoi tessellation partitions the dataset into subvolumes for precise spatial and chemical analysis. This methodology is also used to extract clusters and dislocations in APT data. While Felfer et al. used DCOM to segment out microstructural features, Peng et al. [103] used clustering algorithms like DBSCAN and KNN to extract solute atoms segregating at an interface. Following this step, a triangular grid is fitted on the selected atoms therefore geometrically describing the interface. Voronoi tessellation is used to select atoms across the fitted grid to evaluate thickness, composition profiles, and interfacial excess.

Wei et al. [104] and Zhou et al. [105] performed automated the identification and quantification of chemical features of GBs that provide quantitative, unbiased, and automated access to GB chemical analyses. In their work, Wei et al. employed supervised machine learning techniques, specifically boosting algorithms, to achieve near-atomic resolution in characterizing the GBs geometrically and compositionally. This approach captured the local change in density of hits on the detector to identify the GB. However, it requires descriptive crystallographic data about the adjacent grains to accurately characterize the GBs. Zhou et al. addressed this issue by using the compositional contrast between the GBs and the surrounding matrix phase to identify them. They trained CNNs on the synthetic images to recognize and label features such as grain interiors, GBs, and triple junctions. The trained model is then used to segment out these regions in the experimental data. After extracting the locations of the GBs, the solute distribution along the GB planes is quantified by employing the methodology established by Felfer et al. [106]. This method enables the analysis of multiple GBs within a single dataset. However, it requires the GBs to be columnar or that the plane perpendicular to the GBs is parallel to the axis of the reconstructed APT data. To address this limitation, Saxena et al. [107] generalized the segmentation of various geometrically complex microstructural features, including GBs through their chemical fingerprint. After segmentation geometry-specific algorithms were used to quantify composition and geometric aspects of the microstructure. The approach is discussed in the next section in more detail.

### 3.4.2. Chemical domain/ Phase segmentation

Quantification of the composition, volume fraction, size, geometry, and spatial distribution of microstructural features is crucial for developing new materials and advancing our understanding of diverse physical phenomena. This requires segmentation of the APT data, which has often been done via voxelisation and isosurfaces. Although isosurfaces can quantitatively describe composition aspects, they only provide a qualitative understanding of the geometry and distribution of microstructural features. They can also be very sensitive to the user-defined hyperparameters (iso-surface values). To enhance the automation and reliability of feature segmentation, Madireddy et al. [108] employed deep learning–based edge detection to capture the interface between the matrix and the precipitate phase. However, limited progress has been made in addressing more complex microstructural features and quantifying their shape and geometric characteristics. Zhou et al.'s CNN-based approach as discussed above can automatically find GBs and dissect complex networks of GBs into simpler planar regions for further analysis, however their approach is only fit for columnar grains [105].

Saxena et al. [107] developed workflows for segmenting and quantifying 3D microstructures in APT datasets in a robust and reproducible manner. The initial step involves segmenting areas with similar chemical compositions, termed chemical domains, notably thermodynamic phases as well as segregation zones. This segmentation is achieved by a multi-stage unsupervised machine learning approach, as shown in Fig. 7. The APT dataset is divided into uniformly sized voxels to gather local composition information (Fig. 7b). The set of all local compositions forms a structured point cloud in compositional space. A Gaussian mixture model is employed for clustering distinct phases in the composition space, and a density-based clustering algorithm in real space isolates different microstructural features within a single phase at the voxel resolution (Fig. 7c-e) [93]. Because these are applied onto the voxelized data, their application is substantially faster than doing it on the point cloud directly. These microstructural entities are then examined for their compositional and geometric properties, including orientation, shape, and thickness.

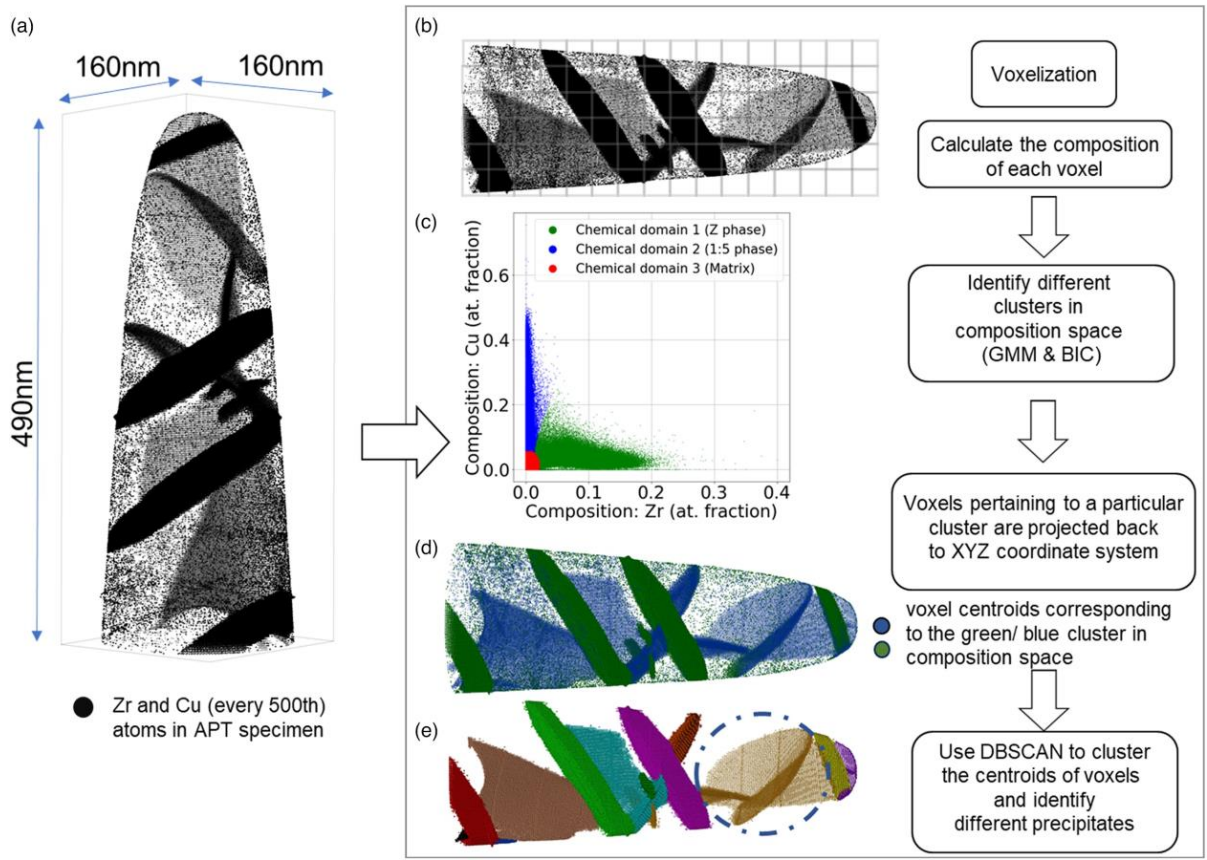

*Fig. 7 Workflow for clustering in composition space-based approach using reconstructed data from a Fe-doped Sm-Co alloy [107].* *(a) APT reconstruction for a Fe-doped Sm–Co alloy. For clarity, only minority elements (Cu, Zr) are shown. (b) Voxelization and per-voxel composition analysis, the gray grid schematically illustrates the voxels. (c) 2D projection of the 5D composition space. Each point corresponds to one voxel. Color coding according to 3 composition clusters identified. (d) Phases in real space at voxel resolution according to composition classification. (e) DBSCAN is used to cluster voxel centroids belonging to each phase to recognize separate*

*individual precipitates. The circled precipitate has a complex morphology. Reprinted with permission from Ref. [107], CC BY 4.0.*

Quantifying 3D microstructures can be challenging due to their intricate geometries. Therefore, in this approach geometry-specific data analysis pipelines were developed. For microstructures containing intertwined plate-like features, a workflow using the U-Net model to simplify the segmented microstructural features into plate-like sub-structures, for example GBs, was developed [109]. This simplification allows for automatic and systematic measurement of in-plane variations in composition and thickness, as well as the relative orientations of the microstructural features. The insights gained from analyzing the in-plane composition and thickness fluctuations are anticipated to enhance our understanding of how the second-phase precipitates influence the magnetic properties of a given Fe-doped Sm-Co alloy, guiding the design of future permanent hard magnets [107].

The approach described in previous study can be extended to extract a wide range of microstructural features and perform both compositional and geometric analysis. While the technique works well for globular and planar features, line-like features—such as segregation at dislocations and stacking faults—often have specialized geometric considerations. These may include branching linear features or dislocations forming loops. To accurately quantify the composition and geometry of such linear features, complex microstructures must be simplified by segmenting individual linear features. Felfer et al. [102] used DCOM to extract dislocations and then computed proxigrams and line-excess plots; however, this method lacks flexibility and automation for more complex dislocation networks. Ghamarian et al. [110] introduced a skeleton-based method to automate dislocation extraction and analysis, but it still lacks algorithms to segment dislocation networks into discrete linear features. In addition, the generated skeleton may not pass through the medial axis of a linear feature, which is crucial for computing accurate composition profiles. Moreover, this method relies on applying a clustering algorithm to atomic positions in the reconstructed data, which can be computationally expensive.

Saxena et al. [111], inspired by the above approaches, developed workflows to segment complex line-like features, particularly dislocations skeletonization [112]. The clustering in composition space approach helps to estimate iso-surface values for constructing iso-composition surfaces. The skeletonization algorithm is then used to reduce the isosurface meshes into a linear graph or skeleton, encapsulating the topological information of the 3D isosurface mesh. The skeleton is used to segment dislocations for detailed composition analysis. Since isosurfaces are computationally cheaper to calculate compared to applying clustering algorithm on the APT point cloud data, this approach is faster than the previous ones. Moreover, advanced algorithms are employed to isolate linear features from intricate microstructures. This allows for the calculation of composition profiles along and perpendicular to each feature, while accounting for local morphological factors such as curvature and cross-sectional shape. Additionally, crystallographic information from APT data is used for orientation analysis, transforming each dislocation segment into the crystal coordinate system. This workflow successfully extracted and analyzed dislocations in a Ni-based alloy, exemplifying its ability to work seamlessly on different APT datasets and its potential to understand the impact of decorated dislocations on material properties.

### 3.5. Standard data analysis workflows

Assuring that the above-mentioned methods deliver reproducible results calls for methods that transform software tools from proof-of-concept to products, which use computing resources efficiently and ease interpretation by humans and machines. Standardization, FAIR principles, and semantic knowledge representation are specific research data management (RDM) methods whose implementation funding agencies are globally pushing for [113] to raise the quality of research in terms of documentation, contextualization, and ultimately foster data reuse using licenses that clarify which portions of research can be reused under which conditions and respecting intellectual properties rights [114]. Standardization thereby improves machine readability for data and metadata to support automation and automated knowledge querying. The overwhelming speed and technical capabilities gained by machine learning methods has left underdeveloped documentation of such workflows beyond architectural contextualization. This is contradictory to the aim that ideally ML methods should be explainable to judge how significant parameterization and algorithmic choices have an effect on the scientific argumentation. We review the work on standardization, workflow modeling, research data management, semantic (web) technologies, and development of software tools that is documented in a region of literature that might often have been too far out on the radar for APT practitioners.

### 3.5.1. Standardization of data analysis workflows to align with the FAIR principles

The FAIR guiding principles for scientific data management and stewardship form a set of strategic considerations, management, and technical suggestions of best practices that envision how infrastructures for the reuse of scholarly data should be developed and offer functionality to support reuse of such data [115]. Introduced via a top-level viewpoint paper [115] with a technical follow-up paper [116] proved effective in driving global acceptance of FAIR as a gold standard that is nowadays in widespread use. The *"FAIR Principles put specific emphasis on enhancing the ability of machines to automatically find and use the data, in addition to supporting its reuse by individuals",* which entangles implementations often with tools from the semantic web technology stack as these enable machines to gain deeper conceptual understanding and reasoning capabilities about data and context. These tools can be applied irrespective of whether ML is used for APT or in any other discipline. Follow-up publications addressed specific aspects of how to develop and align the output of research software (FAIR4RS) [117], artificial intelligence [118], or workflows [119] with the ideas and aims of the FAIR principles, and how to benchmark these tools quantitatively [120].

Materials characterization with APT combines physical experimentation and computational steps into sets of workflows for data reconstruction and processing. Protocol or standard operation procedure are synonyms for workflow especially in experiments. Research on workflows emerged in computer science with nowadays numerous developments and application use cases in the condensed-matter physics and bioscience communities, which have produced several software frameworks and tools for defining, running and documenting workflows. AiiDA [121] for ab-initio modelling and pyiron [122] are a few examples from the hundreds of such tools [123].

In APT, workflows show substantial differences in implementation, complexity, documentation, and granularization of data and metadata. These developments and the narrative of reporting analyses of APT data as formalized and documented workflows has not found widespread use by the community yet, mainly because the analyses still rely on human expertise [124, 125]. None of the existing complementary open-source software tools [126] are frequently used in comparison to commercial software, which has hampered the motivation of substantial efforts to make these tools interoperable with one another via standardized APT computational workflows. Tools for ML in atom probe are no exception [5, 19, 81, 99, 105, 107]. Mismatch in (semantic) concepts, proliferation of jargon, and unclear contextualization are common. APT research is typically locked behind paywalls, metadata of workflows and associated datasets are often neither available as supplementary material nor shared frequently on data repositories.

The situation is improving though with papers and personal exchange at workshops and conferences becoming complemented by digitalization efforts. Miller's reporting [127] methods for associating a mass-to-charge to an elemental or ionic identity is the de facto current standard, even if imperfect and debated [48, 128]. Ceguerra et al. raised the vision with an early demonstrator [129] that it is useful to share results and software tools for APT. This idea was refined at least three times, though always with the important simplification and constraint to work only with those datasets of the authors themselves or their coauthors and not sharing the raw data [7, 130, 131]. The proposal of Ringer was reassessed as a work package by the FAIRmat consortium of the German National Research Data Infrastructure [132]. This has resulted in several key additions to align better with the FAIR principles specifically adding ontologies that represent the majority of data and metadata typically collected in APT, along with several steps of typical computational workflows for post-processing. These ontologies are defined via customizations of the semantic file format NeXus [133, 134] towards an ontology that is currently in an active standardization process by the NeXus International Advisory Board Committee [135], and supported by community feedback [136] to be used as suggested best practice. The ontology is supported with several reference implementations [124, 125, 137] tailored towards use cases in APT via the RDM software NOMAD [138]. These preliminary efforts are necessary to build the data infrastructure for fully deploying ML's capabilities. Furthermore, these efforts can serve as examples how future ML software can also arrive at using equally strong semantic documentation of the configuration and results of ML workflows.

### 3.5.2. Electronic Lab Notebook (ELN)

Capturing the relevant metadata and documenting complete workflows is facilitated by the use of ELNs [139], laboratory information management system (LIMS), and tools or frameworks for RDM [138, 140] that are software products assisting scientists with collecting and managing research data [18, 141]. Many open-source and proprietary products are available that show differences in and overlaps of functionalities and target communities [142]. ELNs focus on customizable templates for structured or unstructured ingestion of metadata and linking of files along a research workflow. ELNs are particularly useful in cases when metadata would otherwise not be digitally stored by the analysis software. Typical examples are sample and specimen preparation in APT or ML workflows that are composed by processing experiments first with e.g. AP Suite, followed by exploratory channeling obtained reconstructions into ML tools orchestrated via Jupyter notebooks and Python. There is potential for improving these tools: For instance functionalities are lacking for keeping track rigorously of the execution flow, and the data is typically accessed via hardcoded paths to files rather than API calls to databases. LIMS focus on functionalities for physical lab resource management such as chemicals in a wet lab. RDM tools typically offer richer functionalities combining services via a database-driven backend for storage and archival with a browser-controlled frontend for data ingestion and exploration, search, visualization, object identification, and permission rights management.

Some advocate that there might not be a need for RDM tools in APT other than the commercial software. However, in materials research, APT is typically combined with other techniques within multi-modal characterization workflows. These include for instance electron microscopy [143], focused-ion beam milling [144], and various spectroscopies [145]. All these can come with their own (often proprietary) solutions. In consequence, data end up disconnected in individual data spaces of several ELN, LIMS, or RDM tools, defeating their purpose somehow. This is where digital identifiers for data, metadata, software tools, and workflows are necessary. Specifically, hashing of digital resources (files, database entries) and version control software with implemented continuous integration workflows for documenting software version are technical solutions to manage research products.

Given that these technological solutions are mature, it is possible to evolve existing ML software prototypes into products such that atom probe scientist can load data and write back results via API calls to commercial software and RDM platforms. Taking advantage of the interconnectivity of data from different techniques is a key argument for setting up dedicated tools, beyond what individual instruments and technology partners can offer. Missing capabilities for provenance tracking so far practically prevent APT groups from effectively managing and citing co-authorships and prevent even a partial reuse of workflows. In effect, groups need to restrict access to many potentially reusable atom probe datasets, analysis scripts and workflows, simply to avoid unintentionally making content publicly available. NOMAD [138] is an open-source RDM solution that has recently been extended to support amongst other condensed-matter physics research communities also atom probe with multi-method data harmonization RDM functionalities. The afore-mentioned NeXus ontologies have been implemented in NOMAD to cover the documentation of data acquisition workflows from instruments powered by AP Suite and custom atom probe instruments, including steps of raw data calibration, reconstruction, and ranging [137, 146]. A plugin infrastructure allows for creating customized deployments and further customization of NOMAD to integrate e.g. with electron microscopy data, the ab-initio computational materials modeling database [132], or the AI toolkit [147].

### 3.5.3. Ontologies: Tools that enable machines to exchange structured and standardized knowledge representations

Connection of research practices can only profit from defining strong semantics and standardized representations. In APT, this should facilitate the deployment of ML, as working with standardized data representations in common ML frameworks can reduce substantial sources of inaccuracies and the typical mandatory data pre-processing. One route forward is to apply and adapt ontologies. These evolved from the historic discipline of philosophy of studying the being towards "representational primitives with which to model a domain of knowledge or discourse" [148]. Different approaches with an increasing semantic strength are distinguished:

1. **List of (technical) terms or social tags (concepts)** are maintained in virtually all scientific domains. Examples in APT are *tip* or *flat-top (microtip coupon).*
2. **Glossary** adds definitions for each concept. Examples are APT-specific terms of ISO 18115-1:2023 [149].
3. **Thesaurus** adds synonyms to concepts (for example, tip is an exact synonym of and jargon term for specimen)

4. **Taxonomy** adds hierarchies via the *is_part_of* relationship. The analysis tree in AP Suite 6 is a typical example of a proprietary taxonomy with point clouds and iso-surfaces as instance data.
5. **Ontology** adds additional types of relationships and constraints on cardinality, existence, or cross-references to concepts from other ontologies.

Several ontologies with respective use cases exist that define concepts within materials science and engineering and condensed-matter physics [150-152]. A comprehensive proposal for an ontology for APT, based on the semantic data schema and file format standardization effort NeXus [134, 135] has been proposed [133, 136] that offers semantic representations tailored for APT-domain-specific documenting of data artifacts and computational steps of data analysis workflows.

**4. Future prospects**

**4.1. Interfacing existing tools with commercial platforms**

Many ML tools for APT, as detailed in Section 3, are currently developed as standalone applications or scripts, requiring users to switch between platforms. Interfacing these tools into commercial software would lower adoption barriers, enhance workflow efficiency, and ensure reproducibility by embedding ML algorithms into standardized analysis pipelines. For instance, incorporating ML-based peak identification algorithms (e.g., ML-ToF) into commercial mass spectrometry tools could significantly automate and standardize peak ranging and elemental identification.

However, this interfacing presents several challenges. First, handling 3D point cloud data with mass-to-charge information demands significant memory and computational power. Resource constraints on commercial platforms may necessitate optimization or hardware acceleration, such as GPU support. Certain preprocessing steps may demand high-performance servers; for example, generating millions of experimental SDMs in ML-APT for analyzing CSRO heavily relies on powerful computing clusters. Effectively coupling these computational resources with commercial platforms will be critical to ensuring seamless and efficient interfacing. Second, selecting appropriate ML model parameters, such as learning rates and network configurations, can be challenging for APT users without ML expertise. To address this, user-friendly interfaces should offer guided parameter selection, automated optimization, and built-in recommendations tailored to dataset characteristics. Simplified quality metrics, interactive visual feedback, and clear documentation are essential to help users understand and adjust model behavior, ensuring accessibility without requiring deep ML knowledge.

### 4.2. Machine learning for APT data rectification

The anisotropic spatial resolution of APT data limits possibilities realizing true atomic-scale analytical tomography [153]. Pioneering efforts by Camus et al. [154], Vurpillot et al. [155], and Moody et al. [1, 156] have tried to use lattice rectification techniques to facilitate the bridging to atomistic simulations in e.g. aluminum alloys. These approaches focus on slightly adjusting atomic positions to align them precisely with their most probable atomic planes in specific crystallographic directions. However, this requires data along three independent crystallographic directions, which is particularly challenging for compositionally-complex materials.

Although fully correcting the 3D lattice in a single step remains a challenging task due to the complexity of atomic displacements and distortions in APT datasets, a step-by-step correction process offers a promising and more feasible alternative. This approach could be especially valuable when working with high-dimensional data, where traditional correction methods may struggle to handle the intricacies of local variations. For instance, diffusion models [157-159], known for their ability to model probabilistic processes, could be employed to treat distorted APT datasets as "noisy" versions of ideal lattice structures. These models would be trained on both synthetic and experimental datasets with known reference structures, enabling them to iteratively remove distortions and realign atoms to their most likely positions within the ideal lattice framework. By applying these models, we may progressively refine the lattice structure, thus minimizing errors and improving the accuracy of the reconstructed atomic configuration.

While the overall process of lattice correction remains complex, certain regions of the 3D lattice may exhibit relatively high crystallographic quality, particularly along specific crystallographic directions (as seen in Supplementary Fig. 1). These high-quality regions can serve as an important asset in the reconstruction process. Deep learning techniques could be employed to leverage these limited high-quality thin 3D zones to enhance the understanding and interpretation of lower-quality regions. By doing so, deep learning algorithms could fill in missing information and refine the overall 3D reconstruction, ultimately leading to a more accurate and high-resolution lattice map. This approach shares similarities with workflows used in transmission electron microscopy (TEM) [160], where advanced image processing and reconstruction techniques are applied to low-resolution TEM data, helping to recover high-resolution structural information. The principles behind these TEM-based workflows could offer valuable guidance and insights for developing robust methods in APT, enabling improved data interpretation and enhancing the accuracy of atomic-scale reconstructions.

### 4.3. Machine learning enhanced APT simulation

For more accurate APT simulations, it is essential to achieve a full-dynamic description of the field evaporation process with "ab-initio" precision, capturing the competition between interatomic and electrostatic forces. This can be accomplished by incorporating field evaporation into molecular dynamics (MD) simulations, as demonstrated in the TAPSim-MD approach [161, 162]. Machine Learning is poised to play a critical role in this context, particularly in accelerating the development and training of interatomic potentials. Once these ML-enhanced potentials are trained, they can be integrated into TAPSim-MD simulations, significantly enhancing both efficiency and predictive accuracy. This integration enables the precise modeling of complex atomic systems involved in APT experiments. By reducing computational costs, ML facilitates faster simulation iterations, leading to deeper insights into field evaporation dynamics and enabling more accurate predictions of atomistic behavior.

Moreover, the concept of digital twinning [163, 164] holds significant potential in the field of APT, offering a means to bridge experimental and simulation efforts. A digital twin is a virtual model that mimics the behavior of a physical system, and in the case of APT, it could be used to simulate the specimen geometry and behavior during experiments. By integrating real-time data from ion trajectory simulations and experimental outputs, digital twins could dynamically model how the tip evolves during the course of an APT experiment, accounting for changes in the shape and size of the tip as atoms are evaporated. This could enable real-time adjustments to simulation parameters, improving the precision of the results. For digital twins to be effective, documentation and standardization should not be treated as an afterthought but rather as fundamental components.

Given the promising potential of machine learning and digital twinning, these technologies are poised to revolutionize APT simulations and experiments. However, these concepts still require further exploration and development to realize their full potential. Moving forward, integrating these advanced computational techniques with the experimental capabilities of APT will be key to driving new innovations and pushing the boundaries of atomic-scale characterization.

## 5. Concluding remarks

In conclusion, APT has proven to be a powerful tool for achieving near-atomic scale 3D compositional mapping. However, its reliance on user expertise for data analysis, coupled with insufficient documentation, has created challenges in standardization and the consistent application of FAIR data principles. Integrating machine learning within a standardized workflow presents a transformative opportunity to address these challenges by enhancing data analysis, reducing user dependency, and improving reproducibility and accuracy. Through the automation of workflows and the provision of deeper insights into material behavior, ML is driving advancements in APT data interpretation. As the field progresses, the continued integration of ML techniques will enhance both the quality of APT data and the speed of simulations, ultimately unlocking new avenues for discovery. To fully realize the potential of machine learning in APT, future efforts should focus on utilizing existing methods, improving data quality, and accelerating simulations to ensure that APT remains at the forefront of atomic-scale characterization.

**Acknowledgements**

YL acknowledges funding by the Alexander von Humboldt Foundation, and the financial support from Deutsche Forschungsgemeinschaft (DFG) under project B04 of the collaborative research center CRC 1625. YL, CF and BG are grateful for financial support from BiGmax, the Max Planck Society's Research Network on Big-Data-Driven Materials Science. MK acknowledges the funding by the DFG under the German National Research Data Infrastructure - project number 460197019 (FAIRmat). BG is grateful for funding by the DFG via the CRC TR 270 project Z01 and the award of the Leibniz Prize 2020. A.S. appreciates funding by Helmholtz School for Data Science in Life, Earth and Energy (HDS-LEE).

**Declaration of Competing Interest**

The authors declare that they have no known competing financial interests or personal relationships that could have appeared to influence the work reported in this paper. MK is task leader for atom probe in the FAIRmat consortium of the German National Research Data Infrastructure (NFDI). FAIRmat extends NOMAD by domain-specific tools for research data management and analysis.

**CRediT authorship contribution statement**

**Yue Li**: Conceptualization, Writing – original draft, Writing – review & editing, Supervision. **Ye Wei**: Writing – original draft, Writing – review & editing. **Alaukik Saxena**: Writing – original draft, Writing – review & editing. **Markus Kühbach**: Writing – original draft, Writing – review & editing. **Christoph Freysoldt**: Writing – review & editing, Supervision. **Baptiste Gault**: Conceptualization, Writing – review & editing, Supervision

**References**

[1] Moody MP, Ceguerra AV, Breen AJ, Cui XY, Gault B, Stephenson LT, et al. Atomically resolved tomography to directly inform simulations for structure–property relationships. Nat Commun. 2014;5:5501.
[2] Gault B, Moody MP, Cairney JM, Ringer SP. Atom probe microscopy: Springer; 2012.
[3] Gault B, Chiaramonti A, Cojocaru-Mirédin O, Stender P, Dubosq R, Freysoldt C, et al. Atom probe tomography. Nat Rev Methods Primers. 2021;1:51.
[4] Haley D, London AJ, Moody MP. Processing APT Spectral Backgrounds for Improved Quantification. Microsc Microanal. 2020;26:964-77.
[5] Li Y, Zhou X, Colnaghi T, Wei Y, Marek A, Li H, et al. Convolutional neural network-assisted recognition of nanoscale L12 ordered structures in face-centred cubic alloys. npj Comput Mater. 2021;7:8.
[6] Gault B, Klaes B, Morgado FF, Freysoldt C, Li Y, De Geuser F, et al. Reflections on the spatial performance of atom probe tomography in the analysis of atomic neighbourhoods. Microsc Microanal. 2022;28:1116-26.
[7] Meier MS, Bagot PAJ, Moody MP, Haley D. Large-Scale Atom Probe Tomography Data Mining: Methods and Application to Inform Hydrogen Behavior. Microsc Microanal. 2023;29:879-89.
[8] Hono K. Nanoscale microstructural analysis of metallic materials by atom probe field ion microscopy. Prog Mater Sci. 2002;47:621-729.
[9] Waugh A, Boyes E, Southon M. Investigations of field evaporation with a field-desorption microscope. Surf Sci. 1976;61:109-42.
[10] Gault B, Vurpillot F, Vella A, Gilbert M, Menand A, Blavette D, et al. Design of a femtosecond laser assisted tomographic atom probe. Rev Sci Instrum. 2006;77.

[11] Geiser BP, Larson DJ, Oltman E, Gerstl S, Reinhard D, Kelly TF, et al. Wide-Field-of-View Atom Probe Reconstruction. Microsc Microanal. 2009;15:292-3.
[12] De Geuser F, Gault B. Reflections on the Projection of Ions in Atom Probe Tomography. Microsc Microanal. 2017;23:238-46.
[13] Lefebvre W, Vurpillot F, Sauvage X. Atom probe tomography: put theory into practice: Academic Press; 2016.
[14] Devaraj A, Perea DE, Liu J, Gordon LM, Prosa TJ, Parikh P, et al. Three-dimensional nanoscale characterisation of materials by atom probe tomography. Int Mater Rev. 2018;63:68-101.
[15] Miller MK, Forbes RG, Miller MK, Forbes RG. The local electrode atom probe: Springer; 2014.
[16] Kelly TF, Miller MK. Atom probe tomography. Rev Sci Instrum. 2007;78.
[17] Gault B. A brief overview of atom probe tomography research. Appl Microsc. 2016;46:117-26.
[18] Bauer S, Benner P, Bereau T, Blum V, Boley M, Carbogno C, et al. Roadmap on data-centric materials science. Modell Simul Mater Sci Eng. 2024;32:063301.
[19] Wei Y, Varanasi RS, Schwarz T, Gomell L, Zhao H, Larson DJ, et al. Machine-learning-enhanced time-of-flight mass spectrometry analysis. Patterns. 2021:100192.
[20] Exertier F, La Fontaine A, Corcoran C, Piazolo S, Belousova E, Peng Z, et al. Atom probe tomography analysis of the reference zircon gj-1: An interlaboratory study. Chem Geol. 2018;495:27-35.
[21] Ulfig R, Kelly TF, Gault B. Promoting Standards in Quantitative Atom Probe Tomography Analysis. Microsc Microanal. 2009;15:260-1.
[22] Gault B, Moody MP, Cairney JM, Ringer SP. Atom probe crystallography. Mater Today. 2012;15:378-86.
[23] Vurpillot F, Da Costa G, Menand A, Blavette D. Structural analyses in three‐dimensional atom probe: a Fourier transform approach. J Microsc. 2001;203:295-302.
[24] Yao L, Moody M, Cairney J, Haley D, Ceguerra A, Zhu C, et al. Crystallographic structural analysis in atom probe microscopy via 3D Hough transformation. Ultramicroscopy. 2011;111:458-63.
[25] Araullo-Peters VJ, Breen A, Ceguerra AV, Gault B, Ringer SP, Cairney JM. A new systematic framework for crystallographic analysis of atom probe data. Ultramicroscopy. 2015;154:7-14.
[26] Haley D, Petersen T, Barton G, Ringer SP. Influence of field evaporation on Radial Distribution Functions in Atom Probe Tomography. Philos Mag. 2009;89:925-43.
[27] Geiser BP, Kelly TF, Larson DJ, Schneir J, Roberts JP. Spatial Distribution Maps for Atom Probe Tomography. Microsc Microanal. 2007;13:437-47.
[28] Moody MP, Gault B, Stephenson LT, Haley D, Ringer SP. Qualification of the tomographic reconstruction in atom probe by advanced spatial distribution map techniques. Ultramicroscopy. 2009;109:815-24.
[29] Gault B, Moody MP, De Geuser F, Tsafnat G, La Fontaine A, Stephenson LT, et al. Advances in the calibration of atom probe tomographic reconstruction. J Appl Phys. 2009;105:034913.
[30] Gault B, Geuser F, Stephenson L, Moody M, Muddle B, Ringer S. Estimation of the Reconstruction Parameters for Atom Probe Tomography. Microsc Microanal. 2008;14:296-305.
[31] Ma K, Blackburn T, Magnussen JP, Kerbstadt M, Ferreirós PA, Pinomaa T, et al. Chromium-based bcc-superalloys strengthened by iron supplements. Acta Mater. 2023;257:119183.

[32] Gault B, Cui XY, Moody MP, De Geuser F, Sigli C, Ringer SP, et al. Atom probe microscopy investigation of Mg site occupancy within δ′ precipitates in an Al–Mg–Li alloy. Scripta Mater. 2012;66:903-6.
[33] Buttard M, Freixes ML, Josserond C, Donnadieu P, Chéhab B, Blandin J-J, et al. Ageing response and strengthening mechanisms in a new Al-Mn-Ni-Cu-Zr alloy designed for laser powder bed fusion. Acta Mater. 2023;259:119271.
[34] Sha G, Yao L, Liao X, Ringer SP, Chao Duan Z, Langdon TG. Segregation of solute elements at grain boundaries in an ultrafine grained Al–Zn–Mg–Cu alloy. Ultramicroscopy. 2011;111:500-5.
[35] Philippe T, De Geuser F, Duguay S, Lefebvre W, Cojocaru-Mirédin O, Da Costa G, et al. Clustering and nearest neighbour distances in atom-probe tomography. Ultramicroscopy. 2009;109:1304-9.
[36] Dumitraschkewitz P, Gerstl SSA, Stephenson LT, Uggowitzer PJ, Pogatscher S. Clustering in Age-Hardenable Aluminum Alloys. Adv Eng Mater. 2018;20:1800255.
[37] Marquis EA, Hyde JM. Applications of atom-probe tomography to the characterisation of solute behaviours. Mater Sci Eng R Rep. 2010;69:37-62.
[38] Sudbrack CK, Noebe RD, Seidman DN. Direct observations of nucleation in a nondilute multicomponent alloy. Phys Rev B. 2006;73:212101.
[39] Zhao H, Gault B, Ponge D, Raabe D, De Geuser F. Parameter free quantitative analysis of atom probe data by correlation functions: Application to the precipitation in Al-Zn-Mg-Cu. Scripta Mater. 2018;154:106-10.
[40] Couturier L, De Geuser F, Deschamps A. Direct comparison of Fe-Cr unmixing characterization by atom probe tomography and small angle scattering. Mater Charact. 2016;121:61-7.
[41] Hellman OC, Vandenbroucke JA, Rüsing J, Isheim D, Seidman DN. Analysis of three-dimensional atom-probe data by the proximity histogram. Microsc Microanal. 2000;6:437-44.
[42] Marceau RKW, Stephenson LT, Hutchinson CR, Ringer SP. Quantitative atom probe analysis of nanostructure containing clusters and precipitates with multiple length scales. Ultramicroscopy. 2011;111:738-42.
[43] Stephenson LT, Moody MP, Gault B, Ringer SP. Nearest neighbour diagnostic statistics on the accuracy of APT solute cluster characterisation. Philos Mag. 2013;93:975-89.
[44] Hornbuckle BC, Kapoor M, Thompson GB. A procedure to create isoconcentration surfaces in low-chemical-partitioning, high-solute alloys. Ultramicroscopy. 2015;159:346-53.
[45] Vurpillot F, Bostel A, Cadel E, Blavette D. The spatial resolution of 3D atom probe in the investigation of single-phase materials. Ultramicroscopy. 2000;84:213-24.
[46] Marceau RKW, Ceguerra AV, Breen AJ, Palm M, Stein F, Ringer SP, et al. Atom probe tomography investigation of heterogeneous short-range ordering in the 'komplex' phase state (K-state) of Fe–18Al (at.%). Intermetallics. 2015;64:23-31.
[47] Dong Y, Etienne A, Frolov A, Fedotova S, Fujii K, Fukuya K, et al. Atom Probe Tomography Interlaboratory Study on Clustering Analysis in Experimental Data Using the Maximum Separation Distance Approach. Microsc Microanal. 2019;25:356-66.
[48] Haley D, Choi P, Raabe D. Guided mass spectrum labelling in atom probe tomography. Ultramicroscopy. 2015;159:338-45.
[49] Guo Y, Wang H, Hu Q, Liu H, Liu L, Bennamoun M. Deep learning for 3d point clouds: A survey. IEEE Trans Pattern Anal Mach Intell. 2020;43:4338-64.

[50] Jordan MI, Mitchell TM. Machine learning: Trends, perspectives, and prospects. Science. 2015;349:255-60.
[51] Saraswat P. Supervised Machine Learning Algorithm: A Review of Classification Techniques. Cham: Springer International Publishing; 2022. p. 477-82.
[52] Wu X, Kumar V, Ross Quinlan J, Ghosh J, Yang Q, Motoda H, et al. Top 10 algorithms in data mining. Knowl Inf Syst. 2008;14:1-37.
[53] Gilpin LH, Bau D, Yuan BZ, Bajwa A, Specter M, Kagal L. Explaining explanations: An overview of interpretability of machine learning. 2018 IEEE 5th International Conference on data science and advanced analytics (DSAA): IEEE; 2018. p. 80-9.
[54] Ke G, Meng Q, Finley T, Wang T, Chen W, Ma W, et al. LightGBM: a highly efficient gradient boosting decision tree. Proceedings of the 31st International Conference on Neural Information Processing Systems. Long Beach, California, USA: Curran Associates Inc.; 2017. p. 3149–57.
[55] Rumelhart DE, Hinton GE, Williams RJ. Learning representations by back-propagating errors. Nature. 1986;323:533-6.
[56] Hornik K. Approximation capabilities of multilayer feedforward networks. Neural Netw. 1991;4:251-7.
[57] Hinton G, Sejnowski TJ. Unsupervised learning: foundations of neural computation: MIT press; 1999.
[58] Saxena A, Prasad M, Gupta A, Bharill N, Patel OP, Tiwari A, et al. A review of clustering techniques and developments. Neurocomputing. 2017;267:664-81.
[59] Bellman RE, Dreyfus SE. Applied dynamic programming: Princeton university press; 2015.
[60] Van Der Maaten L, Postma EO, Van Den Herik HJ. Dimensionality reduction: A comparative review. J Mach Learn Res. 2009;10:13.
[61] Pu J-M, Chen S, Zhang T-Y. Machine learning assisted crystallographic reconstruction from atom probe tomographic images. J Phys: Condens Matter. 2024;37:035901.
[62] Qi CR, Su H, Mo K, Guibas LJ. Pointnet: Deep learning on point sets for 3d classification and segmentation. Proceedings of the IEEE conference on computer vision and pattern recognition2017. p. 652-60.
[63] Sipiran I, Bustos B. Harris 3D: a robust extension of the Harris operator for interest point detection on 3D meshes. Vis Comput. 2011;27:963-76.
[64] Wolff M, Stephens W. A pulsed mass spectrometer with time dispersion. Rev Sci Instrum. 1953;24:616-7.
[65] Maher S, Jjunju FP, Taylor S. Colloquium: 100 years of mass spectrometry: Perspectives and future trends. Rev Mod Phys. 2015;87:113-35.
[66] Sulzer P, Petersson F, Agarwal B, Becker KH, Jürschik S, Märk TD, et al. Proton transfer reaction mass spectrometry and the unambiguous real-time detection of 2, 4, 6 trinitrotoluene. Anal Chem. 2012;84:4161-6.
[67] Pedersen S, Herek J, Zewail A. The validity of the" diradical" hypothesis: direct femtoscond studies of the transition-state structures. Science. 1994;266:1359-64.
[68] Liebscher CH, Stoffers A, Alam M, Lymperakis L, Cojocaru-Mirédin O, Gault B, et al. Strain-induced asymmetric line segregation at faceted Si grain boundaries. Phys Rev Lett. 2018;121:015702.
[69] Biesinger MC, Paepegaey P-Y, McIntyre NS, Harbottle RR, Petersen NO. Principal component analysis of TOF-SIMS images of organic monolayers. Anal Chem. 2002;74:5711-6.

[70] McCombie G, Staab D, Stoeckli M, Knochenmuss R. Spatial and spectral correlations in MALDI mass spectrometry images by clustering and multivariate analysis. Anal Chem. 2005;77:6118-24.
[71] Bluestein BM, Morrish F, Graham DJ, Guenthoer J, Hockenbery D, Porter PL, et al. An unsupervised MVA method to compare specific regions in human breast tumor tissue samples using ToF-SIMS. Analyst. 2016;141:1947-57.
[72] Verbeeck N, Caprioli RM, Van de Plas R. Unsupervised machine learning for exploratory data analysis in imaging mass spectrometry. Mass Spectrom Rev. 2020;39:245-91.
[73] Vurpillot F, Hatzoglou C, Radiguet B, Da Costa G, Delaroche F, Danoix F. Enhancing element identification by expectation–maximization method in atom probe tomography. Microsc Microanal. 2019;25:367-77.
[74] Mikhalychev A, Vlasenko S, Payne T, Reinhard D, Ulyanenkov A. Bayesian approach to automatic mass-spectrum peak identification in atom probe tomography. Ultramicroscopy. 2020;215:113014.
[75] Waugh AR, Boyes ED, Southon MJ. Investigations of field evaporation with a field-desorption microscope. Surf Sci. 1976;61:109-42.
[76] Vurpillot F, Oberdorfer C. Modeling Atom Probe Tomography: A review. Ultramicroscopy. 2015;159:202-16.
[77] Oberdorfer C, Eich SM, Schmitz G. A full-scale simulation approach for atom probe tomography. Ultramicroscopy. 2013;128:55-67.
[78] Breen AJ, Day AC, Lim B, Davids WJ, Ringer SP. Revealing latent pole and zone line information in atom probe detector maps using crystallographically correlated metrics. Ultramicroscopy. 2023;243:113640.
[79] Wright SI, Adams BL. Automatic analysis of electron backscatter diffraction patterns. Metall Trans A. 1992;23:759-67.
[80] Adams BL, Wright SI, Kunze K. Orientation imaging: The emergence of a new microscopy. Metall Trans A. 1993;24:819-31.
[81] Wei Y, Gault B, Varanasi RS, Raabe D, Herbig M, Breen AJ. Machine-learning-based atom probe crystallographic analysis. Ultramicroscopy. 2018;194:15-24.
[82] Li Y, Wei Y, Wang Z, Liu X, Colnaghi T, Han L, et al. Quantitative three-dimensional imaging of chemical short-range order via machine learning enhanced atom probe tomography. Nat Commun. 2023;14:7410.
[83] Wu Y, Zhang F, Yuan X, Huang H, Wen X, Wang Y, et al. Short-range ordering and its effects on mechanical properties of high-entropy alloys. J Mater Sci Technol. 2021;62:214-20.
[84] Wang J, Jiang P, Yuan F, Wu X. Chemical medium-range order in a medium-entropy alloy. Nat Commun. 2022;13:1021.
[85] Zhang R, Zhao S, Ding J, Chong Y, Jia T, Ophus C, et al. Short-range order and its impact on the CrCoNi medium-entropy alloy. Nature. 2020;581:283-7.
[86] Radecka A, Bagot P, Martin T, Coakley J, Vorontsov V, Moody M, et al. The formation of ordered clusters in Ti–7Al and Ti–6Al–4V. Acta Mater. 2016;112:141-9.
[87] Zhang R, Zhao S, Ophus C, Deng Y, Vachhani SJ, Ozdol B, et al. Direct imaging of short-range order and its impact on deformation in Ti-6Al. Sci Adv. 2019;5:eaax2799.
[88] Martin TL, Radecka A, Sun L, Simm T, Dye D, Perkins K, et al. Insights into microstructural interfaces in aerospace alloys characterised by atom probe tomography. Mater Sci Technol. 2016;32:232-41.

[89] Vaumousse D, Cerezo A, Warren PJ. A procedure for quantification of precipitate microstructures from three-dimensional atom probe data. Ultramicroscopy. 2003;95:215-21.

[90] Stephenson LT, Moody MP, Liddicoat PV, Ringer SP. New Techniques for the Analysis of Fine-Scaled Clustering Phenomena within Atom Probe Tomography (APT) Data. Microsc Microanal. 2007;13:448-63.

[91] Ghamarian I, Marquis EA. Hierarchical density-based cluster analysis framework for atom probe tomography data. Ultramicroscopy. 2019;200:28-38.

[92] Wang J, Schreiber DK, Bailey N, Hosemann P, Toloczko MB. The Application of the OPTICS Algorithm to Cluster Analysis in Atom Probe Tomography Data. Microsc Microanal. 2019;25:338-48.

[93] Zelenty J, Dahl A, Hyde J, Smith GDW, Moody MP. Detecting Clusters in Atom Probe Data with Gaussian Mixture Models. Microsc Microanal. 2017;23:269-78.

[94] Vurpillot F, Da Costa G, Menand A, Blavette D. Structural analyses in three-dimensional atom probe: a Fourier transform approach. J Microsc. 2001;203:295-302.

[95] Marceau RKW, Ceguerra AV, Breen AJ, Raabe D, Ringer SP. Quantitative chemical-structure evaluation using atom probe tomography: Short-range order analysis of Fe–Al. Ultramicroscopy. 2015;157:12-20.

[96] He M, Davids WJ, Breen AJ, Ringer SP. Quantifying short-range order using atom probe tomography. Nat Mater. 2024.

[97] Ceguerra AV, Moody MP, Powles RC, Petersen TC, Marceau RK, Ringer SP. Short-range order in multicomponent materials. Acta Crystallogr Sect A: Found Crystallogr. 2012;68:547-60.

[98] Yan K, Xu Y, Niu J, Wu Y, Li Y, Gault B, et al. Unraveling the origin of local chemical ordering in Fe-based solid-solutions. Acta Mater. 2024;264:119583.

[99] Li Y, Colnaghi T, Gong Y, Zhang H, Yu Y, Wei Y, et al. Machine Learning-Enabled Tomographic Imaging of Chemical Short-Range Atomic Ordering. Adv Mater. 2024;36:2407564.

[100] Rao Z, Li Y, Zhang H, Colnaghi T, Marek A, Rampp M, et al. Direct recognition of crystal structures via three-dimensional convolutional neural networks with high accuracy and tolerance to random displacements and missing atoms. Scripta Mater. 2023;234:115542.

[101] Yu J, Wang Z, Saksena A, Wei S, Wei Y, Colnaghi T, et al. 3D deep learning for enhanced atom probe tomography analysis of nanoscale microstructures. Acta Mater. 2024;278:120280.

[102] Felfer P, Ceguerra A, Ringer S, Cairney J. Applying computational geometry techniques for advanced feature analysis in atom probe data. Ultramicroscopy. 2013;132:100-6.

[103] Peng Z, Lu Y, Hatzoglou C, Kwiatkowski da Silva A, Vurpillot F, Ponge D, et al. An Automated Computational Approach for Complete In-Plane Compositional Interface Analysis by Atom Probe Tomography. Microsc Microanal. 2019;25:389-400.

[104] Wei Y, Peng Z, Kühbach M, Breen A, Legros M, Larranaga M, et al. 3D nanostructural characterisation of grain boundaries in atom probe data utilising machine learning methods. PLOS ONE. 2019;14:e0225041.

[105] Zhou X, Wei Y, Kühbach M, Zhao H, Vogel F, Kamachali RD, et al. Revealing in-plane grain boundary composition features through machine learning from atom probe tomography data. Acta Mater. 2022;226:117633.

[106] Felfer P, Scherrer B, Demeulemeester J, Vandervorst W, Cairney JM. Mapping interfacial excess in atom probe data. Ultramicroscopy. 2015;159:438-44.

[107] Saxena A, Polin N, Kusampudi N, Katnagallu S, Molina-Luna L, Gutfleisch O, et al. A Machine Learning Framework for Quantifying Chemical Segregation and Microstructural Features in Atom Probe Tomography Data. Microsc Microanal. 2023;29:1658-70.

[108] Madireddy S, Chung D-W, Loeffler T, Sankaranarayanan SKRS, Seidman DN, Balaprakash P, et al. Phase Segmentation in Atom-Probe Tomography Using Deep Learning-Based Edge Detection. Sci Rep. 2019;9:20140.
[109] Ronneberger O, Fischer P, Brox T. U-Net: Convolutional Networks for Biomedical Image Segmentation. Cham: Springer International Publishing; 2015. p. 234-41.
[110] Ghamarian I, Yu LJ, Marquis EA. Morphological classification of dense objects in atom probe tomography data. Ultramicroscopy. 2020;215:112996.
[111] Saxena A, Kühbach M, Katnagallu S, Kontis P, Gault B, Freysoldt C. Analyzing Linear Features in Atom Probe Tomography Datasets using Skeletonization. Microsc Microanal. 2024;30.
[112] Tagliasacchi A, Alhashim I, Olson M, Zhang H. Mean Curvature Skeletons. Computer Graphics Forum. 2012;31:1735-44.
[113] Services ECaPE. Cost-benefit analysis for FAIR research data : cost of not having FAIR research data: Publications Office; 2018.
[114] Guldimann P, Spiridonov A, Staab R, Jovanović N, Vero M, Vechev V, et al. COMPL-AI Framework: A Technical Interpretation and LLM Benchmarking Suite for the EU Artificial Intelligence Act. arXiv preprint arXiv:241007959. 2024.
[115] Wilkinson MD, Dumontier M, Aalbersberg IJ, Appleton G, Axton M, Baak A, et al. The FAIR Guiding Principles for scientific data management and stewardship. Sci Data. 2016;3:1-9.
[116] Jacobsen A, de Miranda Azevedo R, Juty N, Batista D, Coles S, Cornet R, et al. FAIR Principles: Interpretations and Implementation Considerations. Data Intell. 2020;2:10–29.
[117] Barker M, Chue Hong NP, Katz DS, Lamprecht A-L, Martinez-Ortiz C, Psomopoulos F, et al. Introducing the FAIR Principles for research software. Sci Data. 2022;9:622.
[118] Huerta EA, Blaiszik B, Brinson LC, Bouchard KE, Diaz D, Doglioni C, et al. FAIR for AI: An interdisciplinary and international community building perspective. Sci Data. 2023;10:487.
[119] Goble C, Cohen-Boulakia S, Soiland-Reyes S, Garijo D, Gil Y, Crusoe MR, et al. FAIR Computational Workflows. Data Intell. 2020;2:108–21.
[120] Rocca-Serra P, Gu W, Ioannidis V, Abbassi-Daloii T, Capella-Gutierrez S, Chandramouliswaran I, et al. The FAIR Cookbook - the essential resource for and by FAIR doers. Sci Data. 2023;10:292.
[121] Huber SP, Zoupanos S, Uhrin M, Talirz L, Kahle L, Häuselmann R, et al. AiiDA 1.0, a scalable computational infrastructure for automated reproducible workflows and data provenance. Sci Data. 2020;7:300.
[122] Janssen J, Surendralal S, Lysogorskiy Y, Todorova M, Hickel T, Drautz R, et al. pyiron: An integrated development environment for computational materials science. Comput Mater Sci. 2019;163:24–36.
[123] Gustafsson OJR, Wilkinson SR, Bacall F, Pireddu L, Soiland-Reyes S, Leo S, et al. WorkflowHub: a registry for computational workflows. arXiv preprint arXiv:241006941. 2024.
[124] Kühbach M, Rielli VV, Primig S, Saxena A, Mayweg D, Jenkins B, et al. On Strong-Scaling and Open-Source Tools for High-Throughput Quantification of Material Point Cloud Data: Composition Gradients, Microstructural Object Reconstruction, and Spatial Correlations. arXiv preprint arXiv:2205.13510; 2022.
[125] Kühbach M, Bajaj P, Zhao H, Çelik MH, Jägle EA, Gault B. On strong-scaling and open-source tools for analyzing atom probe tomography data. npj Comput Mater. 2021;7:21.
[126] Kühbach M, Menon S, Saxena A, Forti M, Kružíková P, Hammerschmidt T, et al. NFDI-Matwerk IUC 09: Infrastructure interfaces with condensed matter physics (collaboration with FAIRmat). Zenodo. 2024:https://doi.org/10.5281/zenodo.12594062.

[127] Miller MK. Proposed XML‑based three‑dimensional atom probe data standard. Surf Interface Anal. 2004;36:601–5.
[128] Hudson D, Smith GDW, Gault B. Optimisation of mass ranging for atom probe microanalysis and application to the corrosion processes in Zr alloys. Ultramicroscopy. 2011;111:480-6.
[129] Ceguerra A, Liddicoat P, Apperley M, Ringer S, Goscinski W. Atom Probe Workbench version 1.0.0: An Australian cloud-based platform for the computational analysis of data from an Atom Probe Microscope (APM), used for chemical and 3D structural materials characterisation at the atomic scale.: University of Sydney; 2014.
[130] Diercks DR, Gorman BP, Gerstl SSA. An Open-Access Atom Probe Tomography Mass Spectrum Database. Microsc Microanal. 2017;23:664–5.
[131] Heller M, Ott B, Dalbauer V, Felfer P. A MATLAB Toolbox for Findable, Accessible, Interoperable, and Reusable Atom Probe Data Science. Microsc Microanal. 2024.
[132] Scheffler M, Aeschlimann M, Albrecht M, Bereau T, Bungartz H-J, Felser C, et al. FAIR data enabling new horizons for materials research. Nature. 2022;604:635–42.
[133] Brockhauser S, Dobener F, Emminger C, Ginzburg L, Hildebrandt R, Mozumder R, et al. nexus-definitions. 2024.
[134] Klosowski P, Koennecke M, Tischler JZ, Osborn R. NeXus: A common format for the exchange of neutron and synchroton data. Phys B: Condens Matter. 1997;241–243:151–3.
[135] Committee TNIAB. The NeXus International Advisory Board Committee. NeXus. 2024.
[136] Gault B, Gong Y-W, Saksena AK, Saksena A, Sauvage X, Bagot PAJ, et al. Towards establishing best practice in the analysis of hydrogen and deuterium by atom probe tomography. Microsc Microanal. 2024;30:1205–20.
[137] Kühbach M, Brockhauser S, Dobener F, Mozumder R, Pielsticker L, Shabih S, et al. pynxtools-apm. 2024.
[138] Scheidgen M, Himanen L, Ladines AN, Sikter D, Nakhaee M, Fekete Á, et al. NOMAD: A distributed web-based platform for managing materials science research data. J Open Source Softw. 2023;8:5388.
[139] Barillari C, Ottoz DSM, Fuentes-Serna JM, Ramakrishnan C, Rinn B, Rudolf F. openBIS ELN-LIMS: an open-source database for academic laboratories. Bioinformatics. 2015;32:638–40.
[140] Liu S, Su Y, Yin H, Zhang D, He J, Huang H, et al. An infrastructure with user-centered presentation data model for integrated management of materials data and services. npj Comput Mater. 2021;7:1–8.
[141] Ward L, Aykol M, Blaiszik B, Foster I, Meredig B, Saal J, et al. Strategies for accelerating the adoption of materials informatics. MRS Bull. 2018;43:683–9.
[142] Higgins SG, Nogiwa-Valdez AA, Stevens MM. Considerations for implementing electronic laboratory notebooks in an academic research environment. Nat Protoc. 2022;17:179–89.
[143] Herbig M. Spatially correlated electron microscopy and atom probe tomography: Current possibilities and future perspectives. Scripta Mater. 2018;148:98-105.
[144] Prosa TJ, Larson DJ. Modern Focused-Ion-Beam-Based Site-Specific Specimen Preparation for Atom Probe Tomography. Microsc Microanal. 2017;23:194-209.
[145] Gault B, Schweinar K, Zhang S, Lahn L, Scheu C, Kim S-H, et al. Correlating atom probe tomography with x-ray and electron spectroscopies to understand microstructure–activity relationships in electrocatalysts. MRS Bull. 2022;47:718-26.

[146] Kühbach M, London AJ, Wang J, Schreiber DK, Mendez Martin F, Ghamarian I, et al. Community-Driven Methods for Open and Reproducible Software Tools for Analyzing Datasets from Atom Probe Microscopy. Microsc Microanal. 2022;28:1038–53.

[147] Sbailò L, Fekete Á, Ghiringhelli LM, Scheffler M. The NOMAD Artificial-Intelligence Toolkit: turning materials-science data into knowledge and understanding. npj Comput Mater. 2022;8:1–7.

[148] Gruber T. Ontology. In: Liu L, ÖZsu MT, editors. Encyclopedia of Database Systems. Boston, MA: Springer US; 2009. p. 1963-5.

[149] ISO 18115-1:2023(en) Surface chemical analysis — Vocabulary — Part 1: General terms and terms used in spectroscopy. 2023.

[150] Valdestilhas A, Bayerlein B, Moreno Torres B, Zia GAJ, Muth T. The Intersection Between Semantic Web and Materials Science. Adv Intell Syst. 2023;5:2300051.

[151] Beygi Nasrabadi H, Norouzi E, Sack H, Skrotzki B. Performance Evaluation of Upper-Level Ontologies in Developing Materials Science Ontologies and Knowledge Graphs. Adv Eng Mater. 2024:2401534.

[152] Bayerlein B, Schilling M, Birkholz H, Jung M, Waitelonis J, Mädler L, et al. PMD Core Ontology: Achieving semantic interoperability in materials science. Mater Des. 2024;237:112603.

[153] Kelly TF. Atomic-Scale Analytical Tomography†. Microsc Microanal. 2017;23:34-45.

[154] Camus PP, Larson DJ, Kelly TF. A method for reconstructing and locating atoms on the crystal lattice in three-dimensional atom probe data. Appl Surf Sci. 1995;87:305-10.

[155] Vurpillot F, Renaud L, Blavette D. A new step towards the lattice reconstruction in 3DAP. Ultramicroscopy. 2003;95:223-9.

[156] Moody MP, Gault B, Stephenson LT, Marceau RK, Powles RC, Ceguerra AV, et al. Lattice rectification in atom probe tomography: Toward true three-dimensional atomic microscopy. Microsc Microanal. 2011;17:226-39.

[157] Ho J, Jain A, Abbeel P. Denoising diffusion probabilistic models. Advances in neural information processing systems. Virtual2020. p. 6840-51.

[158] Phan J, Sarmad M, Ruspini L, Kiss G, Lindseth F. Generating 3D images of material microstructures from a single 2D image: a denoising diffusion approach. Sci Rep. 2024;14:6498.

[159] Xu H, Lei Y, Chen Z, Zhang X, Zhao Y, Wang Y, et al. Bayesian Diffusion Models for 3D Shape Reconstruction. Proceedings of the IEEE/CVF Conference on Computer Vision and Pattern Recognition2024. p. 10628-38.

[160] Han Y, Jang J, Cha E, Lee J, Chung H, Jeong M, et al. Deep learning STEM-EDX tomography of nanocrystals. Nat Mach Intell. 2021;3:267-74.

[161] Qi J, Oberdorfer C, Windl W, Marquis EA. Ab initio simulation of field evaporation. Phys Rev Mater. 2022;6:093602.

[162] Qi J, Oberdorfer C, Marquis EA, Windl W. Origin of enhanced zone lines in field evaporation maps. Scripta Mater. 2023;230:115406.

[163] VanDerHorn E, Mahadevan S. Digital Twin: Generalization, characterization and implementation. Decis Support Syst. 2021;145:113524.

[164] Yavari R, Riensche A, Tekerek E, Jacquemetton L, Halliday H, Vandever M, et al. Digitally twinned additive manufacturing: Detecting flaws in laser powder bed fusion by combining thermal simulations with in-situ meltpool sensor data. Mater Des. 2021;211:110167.

**Supplementary information**

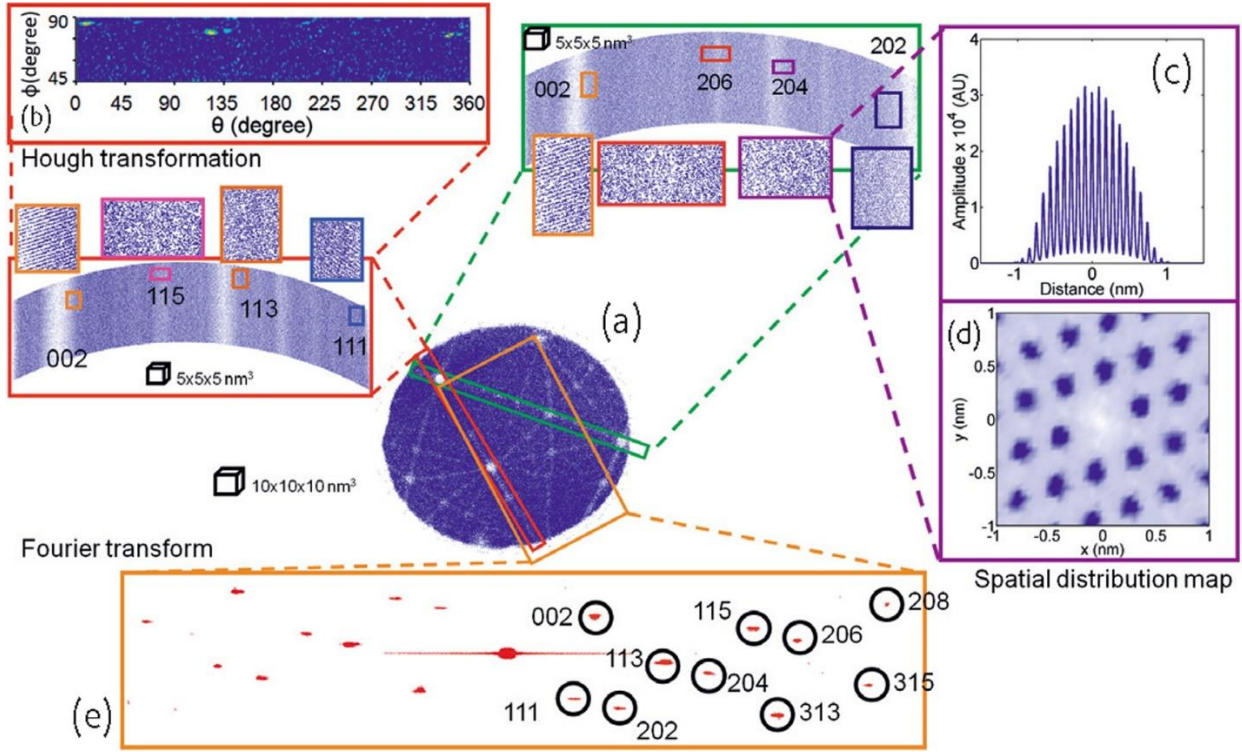

***Supplementary Fig. 1 Periodic feature extraction methods applied to a pure-Al APT dataset.*** *(a) Visualization of the data with close-up views of multiple atomic plane families. (b) Employing Hough transform. (c) and (d) Generating z- and xy-spatial distribution maps, respectively. (e) Using Fourier transform [22]. Reprinted with permission from Ref. [22], CC BY 3.0.*

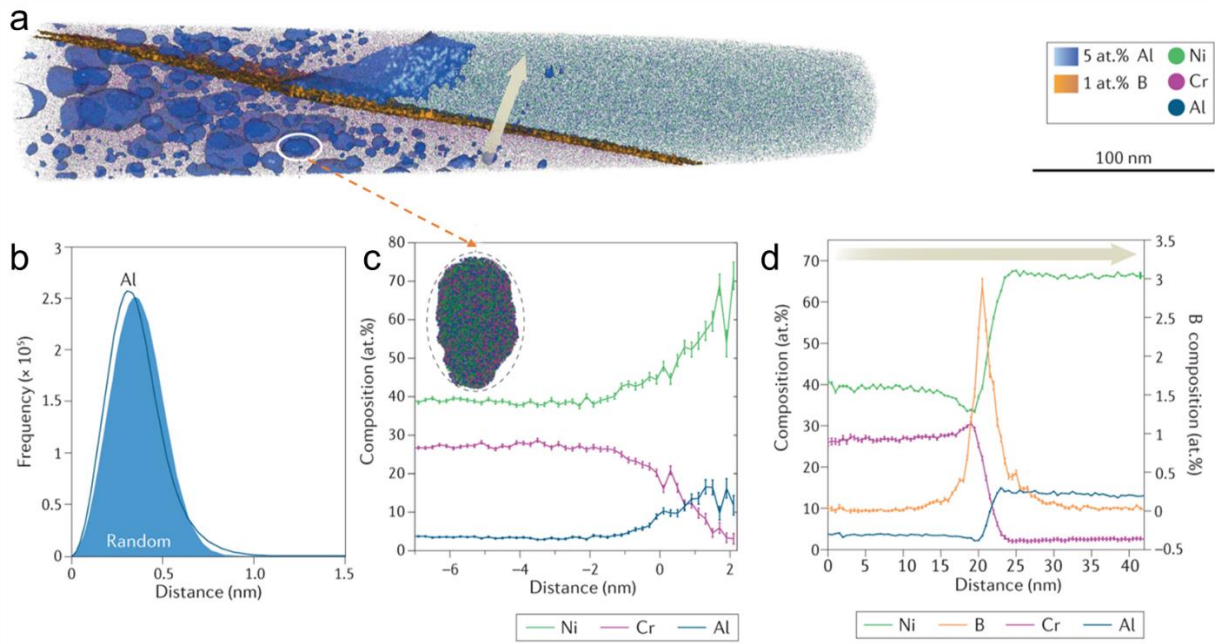

***Supplementary Fig. 2 APT data of an additively manufactured Ni-based alloy** [3]. (a) Point-cloud and iso-concentration surfaces revealing the interfaces between γ and γ' phases (blue) and a grain boundary where B strongly segregates (orange). (b) The 1$^{st}$ NN distribution for Al indicating the clustering/precipitation tendency visible from (a). (c) Composition profile presented as a proximity histogram from the iso-concentration surface delineating the precipitate shown in (a). (d) Composition in a cylinder 20 nm in diameter across the grain boundary calculated along the arrow in (a), enabling quantification of B segregation and γ and γ' phase compositions. Reprinted with permission from Ref. [3].*

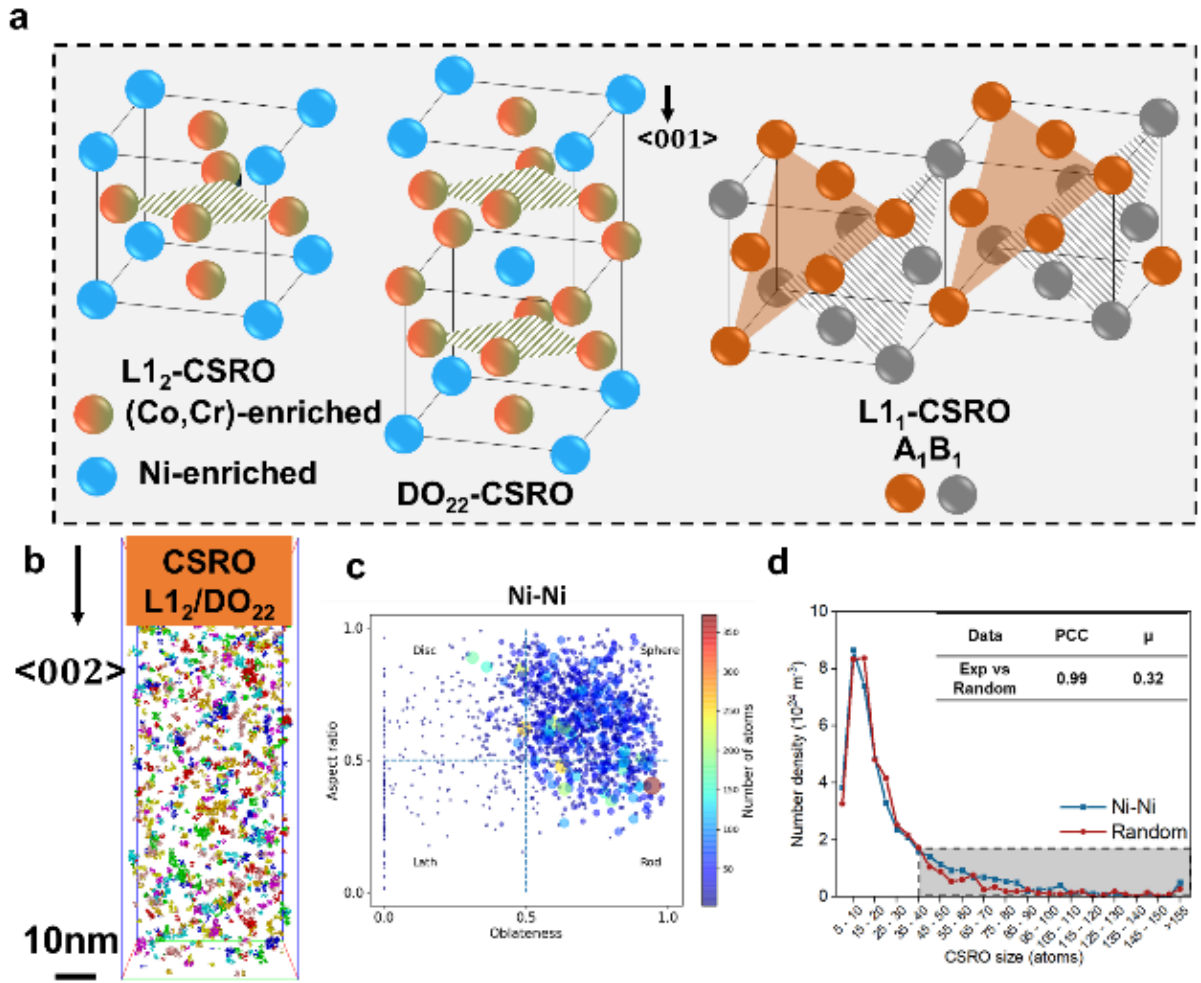

**Supplementary Fig. 3** *Multiple types of CSRO domains in 3D in an annealing CoCrNi alloy obtained by proposed machine-learning APT approach. (a) Experimental derived CSRO configurations and elemental site occupancies. (b) 3D map of the $L1_2/DO_{22}$ type CSROs. (c, d) The corresponding morphology and size distributions [99]. Reprinted with permission from Ref. [99], CC BY 4.0.*